\documentclass[aps,pra,reprint,groupedaddress]{revtex4-1}
\usepackage{graphicx}
\usepackage{xcolor}
\usepackage{amsmath, amssymb}

\begin{document}

\def\om{\omega}
\def\ep{\epsilon} 
\def\epT{\boldsymbol{\epsilon}}
\def\epv{\epsilon_0}
\def\muT{\boldsymbol{\mu}}
\def\muv{\mu_0}
\def\Id{\mathsf{I_d}}
\def\I{\mathsf{I}}
\def\IOmj{\mathsf{I}_{\Omega_j}}
\def\dsum{\displaystyle\sum}
\def\r{\boldsymbol{r}}
\def\k{\boldsymbol{k}}
\def\x{\boldsymbol{x}}
\def\xsf{\textit{\textbf{\textsf{x}}}}
\def\y{\boldsymbol{y}}
\def\z{\boldsymbol{z}}
\def\E{\boldsymbol{E}}
\def\W{\boldsymbol{W}}
\def\P{\boldsymbol{P}}
\def\Esf{\mbox{\sffamily\bfseries\itshape{E}}}
\def\H{\boldsymbol{H}}
\def\Hsf{\mbox{\sffamily\bfseries\itshape{H}}}
\def\D{\boldsymbol{D}}
\def\Ae{\boldsymbol{A}_L}
\def\AD{\boldsymbol{A_D}}
\def\F{\boldsymbol{F}}
\def\dfrac{\displaystyle\frac}
\def\dint{\displaystyle\int}
\def\rot{\nabla\times}
\def\grad{\nabla}
\def\i{\mathrm{i}}
\def\ie{\textit{i.e. }}	
\def\M{\boldsymbol{M}}
\def\dvx{\dfrac{\partial}{\partial x}}
\def\dvy{\dfrac{\partial}{\partial y}}
\def\dvxs{\dfrac{\partial^2}{\partial x^2}}
\def\dvys{\dfrac{\partial^2}{\partial y^2}}

\def\N{\mathbb{N}}
\def\Z{\mathbb{Z}}
\def\R{\mathbb{R}}
\def\C{\mathbb{C}}

\def\epinf{\epsilon_\infty}
\def\epinfi{\epsilon_{\infty_j}}
\def\OmD{\Omega_D}
\def\OmDi{\Omega_{D_i}}
\def\omD{\om_D}
\def\omg{\omega_r}
\def\ompm{\omega_\pm}
\def\omDi{\om_{D_i}}
\def\gammaD{\gamma_D}
\def\gammaDi{\gamma_{D_i}}
\def\OmL{\Omega_L}
\def\OmLi{\Omega_{L_i}}
\def\omL{\om_L}
\def\omij{\om_{D,jk}}
\def\gammaL{\gamma_L}
\def\gammaij{\gamma_{D,jk}} 
\def\Deltaep{\Delta\ep}
\def\Deltaepij{\Delta\ep_{jk}}

%

\title{Calculation and analysis of complex band structure
in dispersive and dissipative two-dimensional photonic crystals.}

\author{Y. Br{\^u}l{\'e}}
\affiliation{CNRS, Aix-Marseille Universit{\'e}, Centrale Marseille, Institut Fresnel UMR 7249, 13013 Marseille, France}
\author{B. Gralak}
\affiliation{CNRS, Aix-Marseille Universit{\'e}, Centrale Marseille, Institut Fresnel UMR 7249, 13013 Marseille, France}
\affiliation{Corresponding author: boris.gralak@fresnel.fr}
\author{G. Dem{\'e}sy}
\affiliation{CNRS, Aix-Marseille Universit{\'e}, Centrale Marseille, Institut Fresnel UMR 7249, 13013 Marseille, France}

\date{\today}

%

\begin{abstract}
Numerical calculation of modes in dispersive and absorptive systems is performed using the finite element method. 
The dispersion is tackled in the frame of an extension of Maxwell's equations where auxiliary fields are 
added to the electromagnetic field. This method is applied to multi-domain cavities and photonic crystals including 
Drude and Drude-Lorentz metals. Numerical results are compared to analytical solutions for simple cavities and 
to previous results of the literature for photonic crystals, showing excellent agreement. The advantages of the 
developed method lie on the versatility of the finite element method regarding geometries, and in sparing 
the use of tedious complex poles research algorithm. Hence the complex spectrum of resonances of 
non-hermitian operators and dissipative systems, like two-dimensional photonic crystal made of absorbing 
Drude metal, can be investigated in detail. The method is used to reveal unexpected features of their complex 
band structures. 
\end{abstract}


\maketitle

\section{Introduction}
Spectral or modal analysis is an essential tool to deal with resonance phenomena in 
optics -- or wave physics in general -- since it provides the natural excitation conditions 
of the system. In optics, metallic structures are of particular interest, since they 
support a vast variety of resonances, resulting from plasmon and geometric structuration. 
These resonances are defined as the eigenfrequencies and corresponding eigenmodes 
associated with the spectral problem of source free Maxwell's equations. 

Metallic structures are classically described by frequency dependent permittivity 
given by Drude or Drude-Lorentz models \cite{jackson_classical_1999}. In this situation
the operator involved in the spectral problem of Maxwell's equations depends on the frequency 
as well as the material dispersion relation. The spectral problem becomes non linear, and 
generally non-hermitian when dissipation takes place. Hence the eigenfrequencies lie in the 
complex plane and the computation of the whole spectrum of resonances is a challenging task.

Numerical calculation of the complex resonances has been performed for two dimensional 
photonic crystals in the cases of square \cite{combes2002} and circular \cite{van2003band} 
rods. These pioneering calculations have been performed using numerical codes solving 
time-harmonic Maxwell's equations combined with a \textit{poles search} algorithm 
in the complex plane, making the tool difficult to use in practice. Hence, new techniques
have been proposed and used to determine the complex resonances. For instance, as for photonic 
crystals, generalizations of the well-known plane wave expansion method 
have been implemented \cite{gu2006applications,alagappan2013optical}. In the case of monotonic 
dispersion relation, \emph{e.g.} away from the materials intrinsic resonances, the cutting plane method 
as proven to be an effective numerical work-around \cite{toader2004photonic,demesy2012solar}.
Finally, finite difference schemes based on the Yee grid have been proposed 
in \cite{ghasemi2012bandgap} or \cite{Fan}. In this last reference, the technique 
is based on the introduction of additional fields leading to an extended 
system where the frequency dispersion is eliminated, which represents a decisive advantage 
for numerical calculations. This framework appears to be a derivation of the auxiliary 
field formalism introduced by A. Tip \cite{tip1997canonical,tip1998linear} in order to 
consider rigorously Maxwell's equations with frequency dispersion and absorption. Indeed, 
in this formalism, auxiliary fields are added to the electromagnetic field to express 
Maxwell's equations with an hermitian operator independent of time. Thus, the auxiliary 
field formalism \cite{tip1997canonical,tip1998linear} is the general frame leading to 
the rigorous treatment of absorption and to significant simplification of frequency dispersion. 

In this article, a version of the auxiliary field formalism is established in order to allow 
the linearization of the spectral problem associated with frequency dispersive materials described 
by Drude or Drude-Lorentz model. A variational form of the resulting augmented system is derived 
and is then implemented into the Finite Element Method (FEM). Next, the presented method is compared 
to semi-analytical results for two dimensional multi-domain closed cavities.
Finally, the method is successfully applied to two-dimensional photonic crystals made of Drude materials and
the fully exhibited richness of the obtained complex spectrum of resonances is discussed. 
\section{The spectral problem}
\subsection{Setup of the problem}
Let $\x,\, \y$ and $\z$ be the unit vectors of the orthogonal Cartesian coordinate system $O_{xyz}$ 
and $(x,y,z)$ the coordinates of the vector $\r$. The time-harmonic regime is considered with a 
$\exp(-\i\om t)$ time dependence, where $\om$ is the possibly complex valued frequency. 
The electromagnetic field is represented by the complex amplitudes $\E$ and $\H$ 
(the time dependency is then omitted). In the case of two-dimensional structures, $\r$ reduces to $(x,y)$ 
and the fields do not depend on the $z$ variable: $\E(\r) = \E(x,y)$ and $\H=\H(x,y)$. Let $\Omega$ denote 
the considered domain, a closed subset of $\mathbb{R}^2$  of boundary $\partial\Omega$. 
The domain $\Omega$ can be constituted of several sub-domains $\Omega_j \in \Omega$ of boundary 
$\partial \Omega_j$ such that $\Omega = \cup \Omega_j$. For each $\Omega_j$, let $\IOmj$ be the 
characteristic function: $\IOmj(\r) = 1 \text{ if } \r \in \Omega_j \text{ and } \IOmj(\r) = 0$ otherwise.
Each sub-domain is made of a linear, possibly graded-index, isotropic, and non-magnetic material. 
The materials are assumed to be electrically frequency dispersive and dissipative. 
The tensor fields of relative complex permittivity $\epT_j$ in each sub-domain $\Omega_j$ are
\begin{equation}
\epT_j(\om) = \epsilon_j(\om)\, \Id \
\end{equation}
and the relative permittivity $\epT$ of the total structure can be written:
\begin{equation}
\epT(\om,\r) = \dsum\limits_{j}\IOmj(\r)\epsilon_j(\om)\, \Id \, .
\end{equation}
It is assumed that the frequency dependence of each $\epsilon_j$ can be modelled by:
\begin{equation}
\ep_j(\om) = \epinfi - \dsum\limits_{k} \dfrac{\om_{p,jk}^2}{\om^2+\i\,\gammaij\,\om - \omij^2} \, ,
\label{EpDL}
\end{equation}
which corresponds to a superposition of Lorentz resonances \cite{jackson_classical_1999}.
This model is known to be particularly relevant to the describe the permittivity of metals in 
the visible range \cite{barchiesi2014errata}.
Note that these functions of $\om$ are analytic in the upper half plane of complex frequencies 
and  hence satisfy the causality requirement. It is stressed that purely dispersive causal systems   
(without absorption) can be tackled by setting the damping constant $\gammaij$ to 0.
Finally, a pure Drude contribution is obtained for $\omij = 0$.

The goal of the present paper is to perform the spectral analysis of the Maxwell's equations
which is based on the calculation of the eigenvalues and the associated eigenvectors of the 
source free  equations:
\begin{equation}
\begin{array}{ll}
        \rot\E(\om,\r)&=\i\om\muv\Id\H(\om,\r),\\
        \rot\H(\om,\r)&=-\i\om\epv\epT(\om,\r)\E(\om,\r)\,,
\end{array}
 \label{Maxwell}
\end{equation}
where $\muv$ and $\epv$ are the vacuum permeability and permittivity.
These equations can be expressed as
\begin{equation}
\M_0(\om,\r) \F_0(\om,\r)=\om \F_0(\om,\r)\, ,
\end{equation}
where 
\begin{equation}
\M_0(\om,\r)=\begin{bmatrix}
   0& \i\,\epT^{-1}(\om,\r)/\epv\rot\\
   -\i/\muv\rot& 0         \\
\end{bmatrix}\,,
\label{EVP}
\end{equation}
and 
\begin{equation}
\F_0(\om,\r)=\begin{bmatrix}
   \E(\om,\r)       \\
   \H(\om,\r)       \\
\end{bmatrix} \, .
\end{equation}
Here $\om$ appears clearly  as the eigenvalue of the operator $\M_0$ with its associated 
eigenvector $\F_0$. As already mentioned, in presence of dispersion, this EigenValue Problem (EVP) 
is non-linear in frequency since $\M_0$ depends on $\om$. When in addition absorption takes place, this 
operator $\M_0$ is also non-hermitian, which requires a numerical computation in the plane of 
complex frequencies. 


\subsection{The auxiliary fields formalism and resonance formalism.}
The auxiliary fields formalism has been introduced and developed by A. Tip since 1997 
\cite{tip1997canonical,tip1998linear}. It is based on the adjunction of auxiliary fields to 
the classical electromagnetic field that leads to the construction of the unique hermitian 
extension \cite{Fig07} of the Maxwell's operator which is linear in frequency. 
In this paper, the purely dispersive case introduced in \cite{GT10} is extended to 
Drude-Lorentz resonances with absorption, see Eq. (\ref{EpDL}). This extended formalism, called 
``resonance'' formalism, is derived hereafter (and in the appendix).
A similar technique has been used in \cite{Fan} to compute the photonic band 
structure of ``metallic'' 2D square rods.

\subsubsection{Single Drude-Lorentz resonance.}
A relative permittivity given by a single Drude-Lorentz resonance is 
considered:
\begin{equation}
\ep(\om) = \epinf - \dfrac{\om_p^2}{\om^2+\i\,\gammaD\,\om - \omD^2} \, .
\label{DrudeLorentzRes}
\end{equation}
Defining 
$\omg = \sqrt{\omD^2-\gammaD^2/4}$, $\ompm = \pm \omg - \i \gammaD/2$, and 
the two following auxiliary fields,
\begin{equation}
\Ae^\pm(\r,t) = \mp \i \, \dfrac{\om_p \,\ompm }{\omg \sqrt{2} } \, 
\dint_{-\infty}^{t}ds\,\exp[-\i\ompm(t-s)]\E(\r,s)\,,
\label{AuxDL}
\end{equation}
Maxwell's equations become in the time regime (see Appendix \ref{AnnexAuxD}):
\begin{equation}
\begin{array}{rl}
\epv\epinf\dfrac{\partial\E}{\partial t}(\r,t) = \hspace*{-2mm} & \rot\H(\r,t) 
-\i\dfrac{\om_p}{\sqrt{2} }\epv \left[\Ae^+(\r,t) + \Ae^-(\r,t)\right]\,,\\[2mm]
\muv\dfrac{\partial\H}{\partial t}(\r,t) = \hspace*{-2mm} & -\rot\E(\r,t)\, ,\\[2mm]
\dfrac{\partial\Ae^\pm}{\partial t}(\r,t) = \hspace*{-2mm} & \mp \i \, 
\dfrac{\om_p\,\ompm}{ \sqrt{2} \omg}\, \E(\r,t) - \i\ompm \Ae^\pm(\r,t) \,.
\end{array}
\label{TempMaxDL}
\end{equation}
As to the harmonic regime, it is deduced from a Fourier decomposition with respect to time:
\begin{equation}
\begin{array}{rl}
-\i\om\epv\epinf\E(\r,\om) = \hspace*{-2mm} & \rot\H(\r,\om) \\[2mm]
& -\i\dfrac{\om_p}{\sqrt{2} } \epv \left[\Ae^+(\r,\om) + \Ae^-(\r,\om)\right]\,,\\[2mm]
-\i\om\muv\H(\r,\om) = \hspace*{-2mm} & -\rot\E(\r,\om)\, ,\\[2mm]
-\i\om\Ae^\pm(\r,\om)  = \hspace*{-2mm} & \mp \i \, \dfrac{\om_p\,\ompm}{ \sqrt{2} \omg} \, \E(\r,\om) 
- \i\ompm \Ae^\pm(\r,\om) \, .
\end{array}
\label{HarmMaxDL}
\end{equation}
This last set of equations can be summarized as 
\begin{equation}
\M(\r)\F(\om,\r)= \om \F(\om,\r)\, ,
\label{EVPAuxDL}
\end{equation}
using the following matrix and vector notations:
\begin{equation}
\M(\r)=\begin{bmatrix}
   0& \dfrac{\i}{\epv\epinf}\rot&\dfrac{\om_p}{ \sqrt{2} \epinf}&\dfrac{\om_p}{\sqrt{2} \epinf }\\
   -\i/\muv\rot& 0   &0&0      \\
\dfrac{\om_p}{ \sqrt{2} \omg} \, \om_+ & 0 & \om_+ & 0\\[1mm]
- \dfrac{\om_p}{ \sqrt{2} \omg} \, \om_- & 0 & 0 & \om_-\\
\end{bmatrix} \, ,
\label{EVPAuxDL}
\end{equation}
and
\begin{equation}
\F(\om,\r)=\begin{bmatrix}
   \E(\om,\r)       \\
   \H(\om,\r)       \\
   \Ae^+(\om,\r)       \\
   \Ae^-(\om,\r)       \\
\end{bmatrix} \, .
\end{equation}
It is important to notice that the operator $\M(\r)$ in Eq. (\ref{EVPAuxDL}) 
is frequency independent.

\subsubsection{Single Drude resonance.}
Let the relative permittivity now be given by the Drude model:
\begin{equation}
\ep(\om) = \epinf - \dfrac{\om_p^2}{\om(\om+\i\,\gammaD)} \, .
\label{Drude}
\end{equation}
In this case, a single auxiliary field is defined as
\begin{equation}
  \AD(\r,t) = -2\i\dfrac{\om_p}{\sqrt{2}}\dint_{-\infty}^{t}ds\,\exp[-\gammaD(t-s)]\E(\r,s)\, .
\label{AuxD}
\end{equation}
Thus, Maxwell's equations in the time regime are
\begin{equation}
\begin{array}{rl}
\epv\epinf\dfrac{\partial\E}{\partial t}(\r,t) = \hspace*{-2mm} & \rot\H(\r,t)
-\i\dfrac{\om_p\epv}{\sqrt{2}}\AD(\r,t) \, ,\\
\muv\dfrac{\partial\H}{\partial t}(\r,t) = \hspace*{-2mm} & -\rot\E(\r,t)\, ,\\
\dfrac{\partial\AD}{\partial t}(\r,t)  = \hspace*{-2mm} & 
-2\i\dfrac{\om_p}{\sqrt{2}}\E(\r,t)-\gammaD\AD(\r,t) \,.
\end{array}
\label{TempMaxD}
\end{equation}
In the time harmonic regime, they become
\begin{equation}
\begin{array}{rl}
-\i\om\epv\epinf\E(\r,\om) =\hspace*{-2mm} & \rot\H(\r,\om)-\i\dfrac{\om_p\epv}{\sqrt{2}}\AD(\r,\om) \, ,\\
-\i\om\muv\H(\r,\om) = \hspace*{-2mm} & -\rot\E(\r,\om)\, ,\\
-\i\om\AD(\r,\om)  = \hspace*{-2mm} & -2\i\dfrac{\om_p}{\sqrt{2}}\E(\r,\om)-\gammaD\AD(\r,\om) \, .
\end{array}
\label{HarmMaxD}
\end{equation}
Finally, a compact expression is obtained as previously using vector and matrix 
notations:
\begin{equation}
\M(\r)\F(\om,\r)=\om \F(\om,\r)\, ,
\label{EVPAuxD}
\end{equation}
where
\begin{equation}
\M(\r)=\begin{bmatrix}
   0& \dfrac{\i}{\epv\epinf}\rot&\dfrac{\om_p}{\sqrt{2}\epinf}\\
   -\dfrac{\i}{\muv}\rot&0&0      \\
2\dfrac{\om_p}{\sqrt{2}}&0&-\i\gammaD\\
\end{bmatrix} \, , 
\end{equation}
and
\begin{equation}
\F(\om,\r)=\begin{bmatrix}
   \E(\om,\r)       \\
   \H(\om,\r)       \\
   \AD(\om,\r)       \\
\end{bmatrix}\,.
\end{equation}

\subsubsection{Superposition of Drude and Drude-Lorentz resonances.}
For the general case of a sum of Drude and Drude-Lorentz resonances as in Eq. (\ref{EpDL}),
the same reasoning leads to the adjunction of a couple of auxiliary fields defined as in 
Eq. (\ref{AuxDL}) for each Drude-Lorentz resonance, and of a single auxiliary field defined as 
in Eq. (\ref{AuxD}) for each Drude resonance.

\subsubsection{Formalism for a mutlidomain structure.}
When several dispersive materials are involved in the total structure $\Omega$, the 
parameters defined in the previous sections (\emph{e.g.} $\epinf,\,\om_p,\,\omg,\,\gammaD$ 
for a single Drude-Lorentz resonance) take different constant values in the different 
subdomains $\Omega_j$. It is then possible to define those parameters (here for example $\epinf$) 
as a function of space:
\begin{equation}
\epinf(\r) =  \dsum\limits_{j}\IOmj(\r)\epinfi
\end{equation}
and so on for all the parameters involved in the chosen permittivity model. 

\subsection{Finite element formulation\label{FEM}}
In this section, the FEM formulation is adapted to the resonance formalism 
in the frame of a single Lorentz resonance. Here, the two auxiliary fields 
$\Ae^+$ and $\Ae^-$ defined in Eq. (\ref{AuxDL}) are introduced. Note that the spatial 
dependency of all the parameters of the Drude-Lorentz resonance in the resonance 
formalism frame is omitted [\emph{e.g.} $\epinf = \epinf(\r)$]. 

The usual treatment of 2D non-conical mount (also called scalar) electromagnetic 
problems, consists in choosing for unknown the out-of-plane component of the electric 
or magnetic fields which is conveniently continuous everywhere. Two linear polarization 
cases of are  traditionally considered, s-polarization [electric field along the z-
direction: $\E = E_z(x,y)\z$] and p-polarization [magnetic field along the z-direction: 
$\H = H_z(x,y)\z$]. Maxwell's equations are combined to form one scalar Helmholtz propagation 
equation involving $E_z$ or $H_z$ solely. 

The generalized Maxwell eigenvector $\F$ in the general case of Eq. (\ref{AuxDL}) contains four 
vector fields belonging to the same function space. However, in two dimensions, the 
seemingly obvious choice of nodal elements for the  discretize the continuous electric 
field $\E = E_z(x,y)\z$ together with edge elements to discretize the in-plane 
discontinuous magnetic field $\H = H_x(x,y)\x+H_y(x,y)\y$ is not appropriate. Indeed, 
$\E$ and $\H$ would not be evaluated at the same points, since the physical quantity
associated to edge elements unknowns is the circulation of the electric field along 
the edges of one mesh triangle, whereas the physical quantity associated to nodal 
elements unknowns is the very value of the field at one node of the mesh. Hence, 
$\E$ and corresponding $\H$ would, in fact, not be evaluated at the same points.
Keeping the full system Eq. (\ref{AuxDL}) would require the notion of dual grid, 
as the interlaced cubic Yee grid \cite{yee1966numerical} for instance, or the use of mixed finite elements.
The choice was made here to work with one element family only by 
classically removing $\H$ from the equation system. However, a 
quadratic EVP is obtained:
\begin{equation}
\begin{array}{ll}
\om^2\epinf\E(\r,\om) =&c^2\rot\rot\E(\r,\om)\\
&+\om\dfrac{\om_p}{\sqrt{2}}\left[\Ae^+(\r,\om)+ \Ae^-(\r,\om)\right]\, ,
\end{array}
\label{HarmHelmDL1}
\end{equation}
\begin{equation}
\om\Ae^+(\r,\om)  = \dfrac{\om_p\,\om_+}{\sqrt{2}\,\omg}\,\E(\r,
\om)+\om_+\,\Ae^+(\r,\om)\, ,
\label{HarmHelmDL2}
\end{equation}
\begin{equation}
\om\Ae^-(\r,\om)  = -\dfrac{\om_p\,\om_-}{\sqrt{2}\,\omg}\,\E(\r,
\om)+\om_-\Ae^-(\r,\om).
\label{HarmHelmDL3}
\end{equation}
This choice is possible thanks to recent advances in linear algebra algorithms that provided 
efficient libraries able to directly tackle such quadratic EVP \cite{slepc-users-manual}.
The weak form of this last equation system is classically obtained by projecting the 
unknown vector fields on weighted vectors fields $\W'$ and integrating over the support 
of the unknowns, $\Omega$ for $\E$ and $\Omega '$ for auxiliary fields $\Ae^+$ and 
$\Ae^{-}$, leading to the following residue:

\begin{equation}
\begin{array}{ll}
\dint_{\Omega} d\Omega&-\om^2\epinf\E\cdot\overline{\W '} + 
c^2\rot\rot\E\cdot\overline{\W '}\\
&+\om\dfrac{\om_p}{\sqrt{2}}\left[\Ae^+ + \Ae^-\right]\cdot\overline{\W '} \\
+ \dint_{\Omega '} d\Omega '& -\om\Ae^+\cdot\overline{\W '} +
\dfrac{\om_p\,\om_+}{\sqrt{2}\,\omg}\,\E\cdot\overline{\W '}\\
& +\om_+\,\Ae^+\cdot\overline{\W '}\\
&-\om\Ae^-\cdot\overline{\W '} -\dfrac{\om_p\,\om_-}{\sqrt{2}\,\omg}\,\E\cdot\overline{\W '}\\
& +\om_-\,\Ae^-\cdot\overline{\W '} \, .
\end{array}
\label{WeakAuxDL}
\end{equation}
Integrating by parts Eq. (\ref{WeakAuxDL}) and using the Green-Ostrogradsky theorem leads to:

\begin{equation}
\begin{array}{ll}
\dint_{\Omega }d\Omega&-\om^2\epinf\E\cdot\overline{\W '} + 
c^2\rot\E\cdot\rot\overline{\W '}\\
&+\om\dfrac{\om_p}{\sqrt{2}}\left[\Ae^+ + \Ae^-\right]\cdot\overline{\W '} \\
+ \dint_{\Omega '} d\Omega '& -\om\Ae^+\cdot\overline{\W '} +
\dfrac{\om_p\,\om_+}{\sqrt{2}\,\omg}\,\E\cdot\overline{\W '}\\
& +\om_+\,\Ae^+\cdot\overline{\W '}\\
&-\om\Ae^-\cdot\overline{\W '} -\dfrac{\om_p\,\om_-}{\sqrt{2}\,\omg}\,\E\cdot\overline{\W '}\\
& +\om_-\,\Ae^-\cdot\overline{\W '}\\
-\dint_{\partial\Omega}dS&(\rot\E \times \bold{n})\cdot\overline{\W '}\, .
\end{array}
\label{IPPWeakAuxDL}
\end{equation}
where $\bold{n}$ refers to the exterior unit vector to the surface $\partial\Omega$ 
enclosing $\Omega$.
The last surface term in Eq. (\ref{IPPWeakAuxDL}) is used to impose boundary conditions 
(Dirichlet, Neumann or Bloch-Floquet conditions \cite{nicolet2004modelling}). Note that 
no particular attention needs to be paid to the auxiliary fields on the boundary of 
$\Omega '$.

As previously mentioned, two cases of polarization are considered (s and p). 
In s-polarization, the electric field and indeed the auxiliary fields have only one 
component along the $O_z$ axis. The problem is scalar and nodal first order elements 
are used as a basis for the unknowns. 
In p-polarization, the fields have two possibly discontinuous components in the plane 
of invariance $O_{xy}$. The use of edge elements (or Whitney forms) is necessary 
\cite{dular1995discrete}. These edge elements naturally take into account the 
discontinuity of the normal component of the electric field across interfaces.
According to the Galerkin formulation, these basis functions are also used as weighted 
functions $\W '$. 

The described weak formulation has been implemented in practice into the FEM freewares 
Gmsh \cite{geuzaine2009gmsh} for mesh generation and visualization, and GetDP 
\cite{dular1998general} as a Finite Element library.

\section{Numerical validation of the method.}

\begin{figure}[h]
\centering
\fbox{\includegraphics[width=0.7\linewidth]{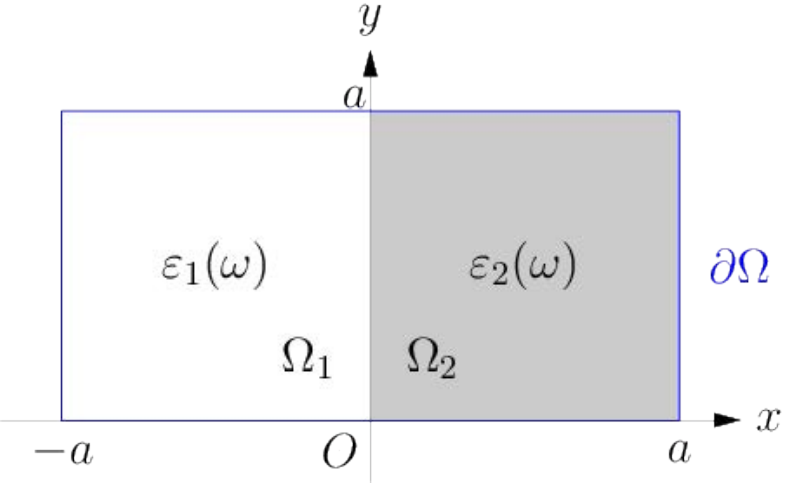}}
\caption{Scheme of the considered closed cavity.}
\label{fig:2Cavite}
\end{figure}

The validity of the method is tested in the case of a two-dimensional closed 
cavity made of two square domains 
$\Omega_1$ and $\Omega_2$, with sides denoted by $a$, and filled with materials of 
relative permittivity $\epsilon_1(\om)$ and 
$\epsilon_2(\om)$ (see Fig. \ref{fig:2Cavite}).
The cavity is delimited by a perfectly conducting metal, with the 
corresponding conditions at the cavity boundary $\partial\Omega$. 

The complex resonances $\om_n$ of this structure can be determined by solving 
the following transcendental equation
defining the dispersion law (see Appendix \ref{EigenCavity} for the derivation 
of the dispersion equation):
\begin{equation}
\dfrac{\tan[\beta_1(\om_n)a]}{\beta_1(\om_n)}+
\dfrac{\tan[\beta_2(\om_n)a]}{\beta_2(\om_n)} = 0
\label{TransS}
\end{equation}
for s-polarization and
\begin{equation}
\dfrac{\beta_1(\om_n)}{\epsilon_1(\om_n)}\tan[\beta_1(\om_n)a]+
\dfrac{\beta_2(\om_n)}{\epsilon_2(\om_n)}\tan[\beta_2(\om_n)a] = 0 \,
\label{TransP}
\end{equation}
for p-polarization. Here the complex functions of $\om$, $\beta_1$ and $\beta_2$, are defined by 
\begin{equation}
\beta_j(\om_n) = \sqrt{\dfrac{\om_n^2}{c^2}\epsilon_j(\om_n)-
\dfrac{q^2\pi^2}{a^2}} \, , \quad j \in \{1,2\}, 
\label{beta}
\end{equation}
where $q \in \mathbb{N} \setminus \{ 0 \}$ for s-polarization and 
$q \in \mathbb{N}$ for p-polarization.

\begin{figure}[h]
\centering
\includegraphics[width=0.80\linewidth]{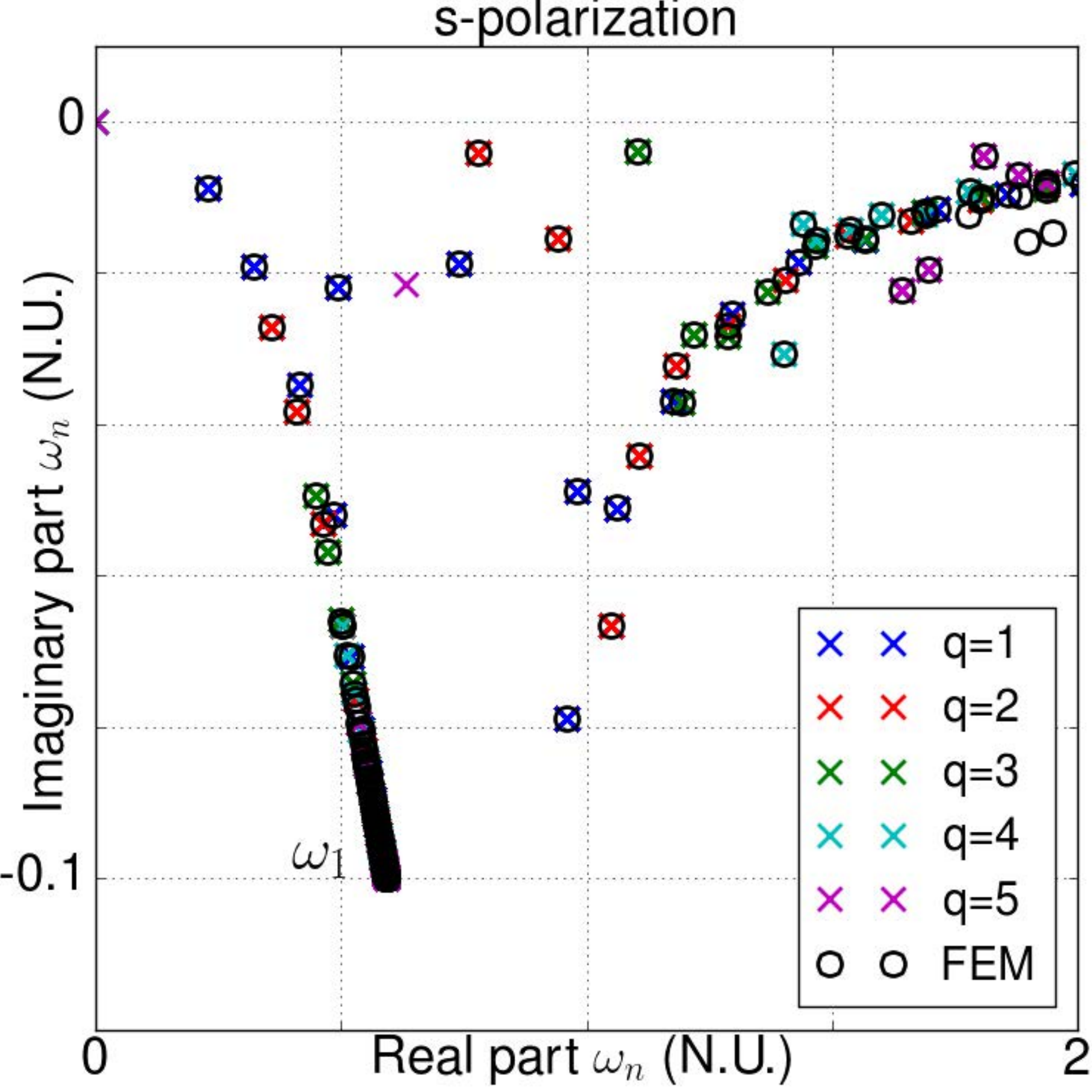}\\[4mm]
\includegraphics[width=0.80\linewidth]{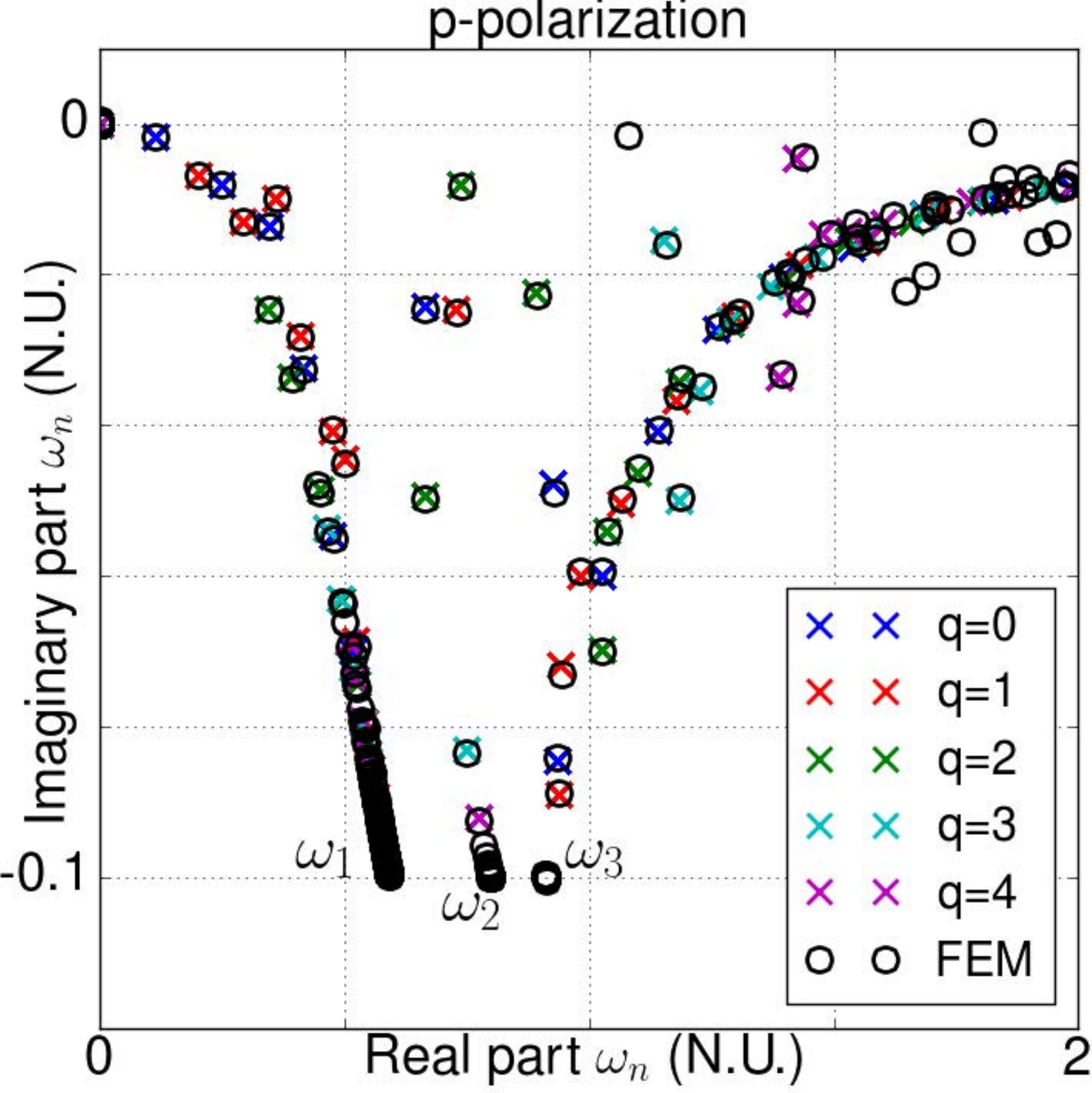}
\caption{Comparison between the results obtained using the FEM method with
auxiliary fields and the semi-analytical method for s- and p-polarization.}
\label{fig:AnaVsFEM}
\end{figure}

\begin{figure}[h]
\centering
\fbox{
\includegraphics[width=0.535\linewidth]{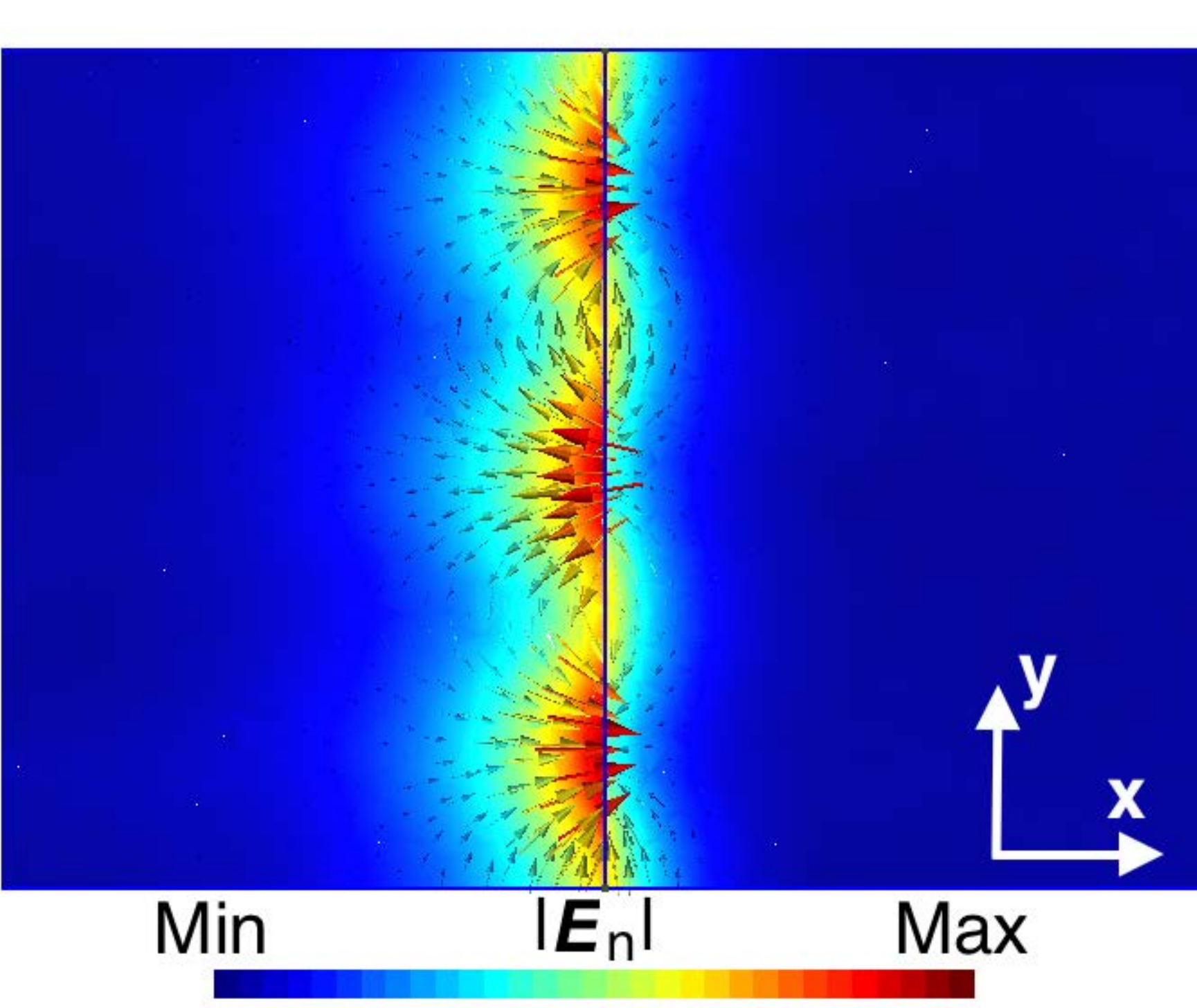}
\includegraphics[width=0.41\linewidth]{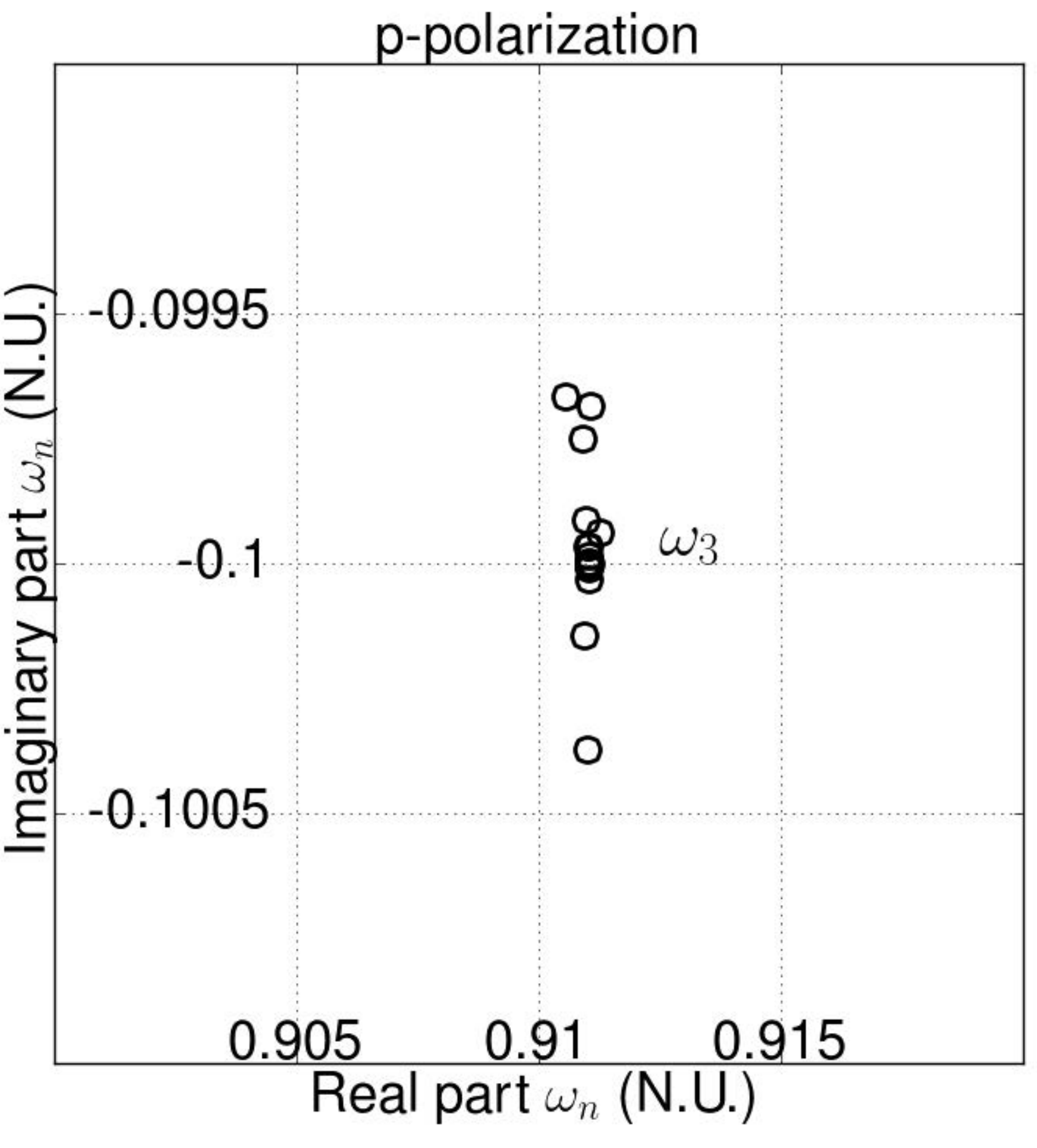}
}
\caption{Left panel: Electric field for one mode of the surface plasmon branch in 
p-polarization at the frequency $\om a / (2 \pi c ) = 0.748-0.083\i$. 
Right panel: Zoom on the spectra of Fig.\ref{fig:AnaVsFEM} around the spurious 
modes produced by the $\ep(\om) = 0$ at $\om = \om_3$.}
\label{fig3}
\end{figure}

The considered cavity has been implemented into the FEM modal method 
described in Sec.~\ref{FEM}.  with the 
permittivities $\ep_1 = 2.0$ (\emph{i.e.} a non dispersive material) and 
\begin{equation}
\ep_2(\om) = \ep_\infty -\dfrac{\omega_p^2 }{ \om^2+ \gamma \i \om - \omega_0^2}\, ,
\label{ep2}
\end{equation}
with
\begin{equation}
\ep_\infty = 3.0 \, , \quad \dfrac{\omega_p a}{2 \pi c} = 1.2 \, , \quad 
\dfrac{\gamma a}{2 \pi c} = 0.2 \, , \quad \dfrac{\omega_0 a}{2 \pi c} = 0.6 \, .
\end{equation}
As the problem is formulated with the electric field, 
Dirichlet boundary conditions are imposed 
on the boundary $\partial\Omega$ of the structure.
Hence, the eigenfrequencies of the structure for both s- and p-polarizations 
are now determined in two different ways: 
using a numerical resolution of the transcendental dispersion 
equations (\ref{TransS}) and (\ref{TransP}) and 
the FEM modal method. The comparison between the results 
of the two methods is shown in Fig.~\ref{fig:AnaVsFEM} for s- (top panel) 
and p-polarization (bottom). 
The results show excellent agreement for s-polarization.
In both s and p polarizations, a first branch of modes is observed to be converging 
towards the (normalized) frequency $(\om_1 a)/(2 \pi c) \simeq 0,59 - 0.1\i$ 
which corresponds to the pole frequency in the expression (\ref{ep2}): 
$|\ep(\om)| \rightarrow \infty$ when $\om \rightarrow \om_1$. It is also 
observed that the spectrum of resonances tends 
to the real axis at high frequencies. Indeed, in this range, 
the value of the permittivity $\ep_2(\om)$ tends to 
$\ep_\infty = 3.0$, and the spectrum of a system with purely real and 
positive permittivity is retrieved. For p-polarization, 
a second branch of modes appears with an accumulation point at the (normalized) 
frequency $(\om_2 a)/(2 \pi c) \simeq 0.8-0.1\i$ where $\ep_2(\om) = - \ep_1$. 
This branch corresponds to the surface plasmon 
modes (see bottom panel in Fig. \ref{fig:AnaVsFEM} and left panel in 
Fig. \ref{fig3}).

For p polarization, spurious modes are also observed around $(\om_3 a)/(2 \pi c) 
\simeq 0.91-0.1\i$ which corresponds 
to the frequency where $\ep_2(\om) =0$. The presence of these spurious modes 
can be explained by the fact that, when $\ep_2(\om)=0$, the divergence free 
condition of the electric field $\nabla\cdot\left[\ep \E\right]=0$ 
fails since $\E$ can take any value in the dispersive media with $\ep_2(\om) =0$.
As a consequence, the precision of the results is 
affected by the presence of these spurious modes 
and the accuracy of the FEM results is degraded in a neighborhood of $\om_3$, 
which probably has overlap with the branch of surface plasmon modes 
around $\om_2$ (cf Fig. \ref{fig3}, right panel).
These spurious modes are not present in s polarization as the electric field 
is oriented along the $O_z$ axis and spatially depends only on $(x,y)$ so 
that the divergence free condition is included in the scalar equation.
These spurious modes can be avoided by the implementation of Lagrangian 
preconditioners into the eigenvalue problem solver.

In addition, it is mentioned that the method has also been validated in the case 
of two dispersive domains with relative permittivities given by a sums of Drude 
and Drude-lorentz resonances. This final test shows that the proposed method can 
be used in practical situations.

The numerical method presented in this paper provides an efficient tool 
to compute and analyze the complex spectrum of dispersive systems, and thus 
it opens new possibilities of investigations in the area of non 
self-adjoint operators which has been rarely explored. The situation of 
a two-dimensional crystal is considered in the next section. 

\section{Complex resonances in a two dimensional photonic crystal}

\begin{figure}[h]
\centering
\fbox{\includegraphics[width=0.97\linewidth]{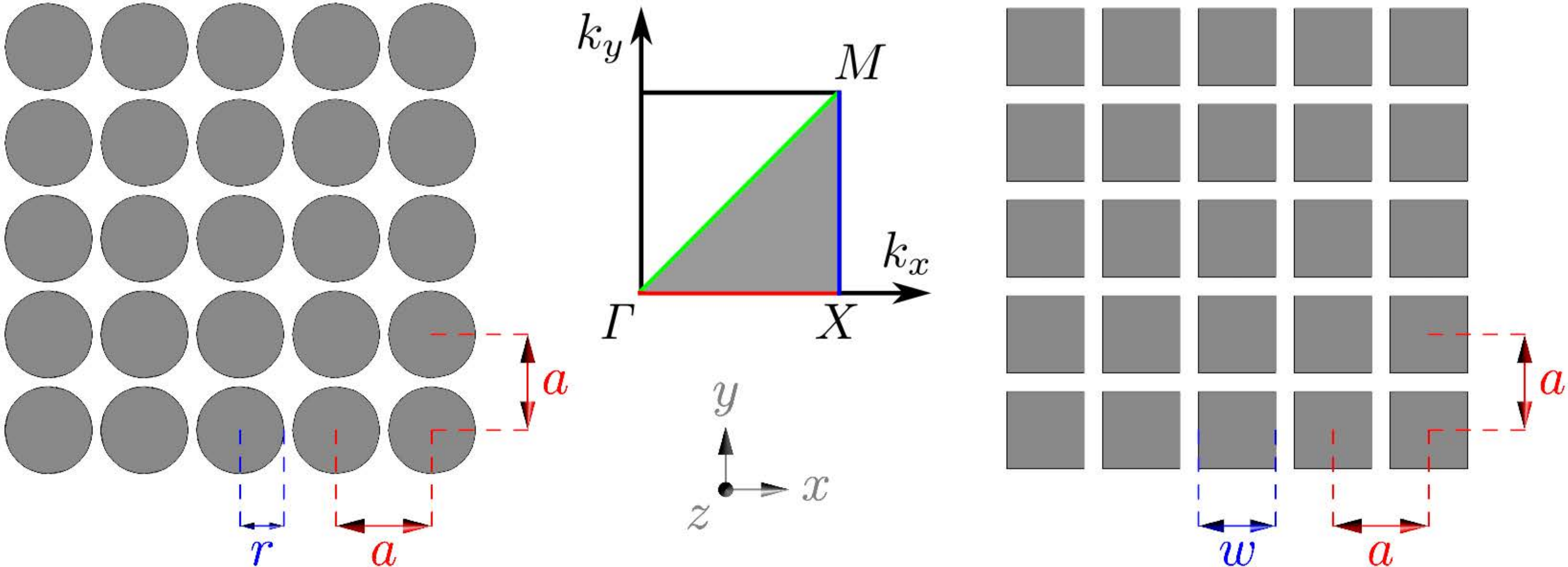}}
\caption{Scheme of the modeled 2D photonic crystals in one invariance plane 
and the associated first reduced Brillouin zone (triangle $\Gamma X M$: 
$\Gamma X$ in red, $\Gamma M$ in green, and $XM$ in  blue).}
\label{fig:SchemeCrystal}
\end{figure}

The two dimensional photonic crystals considered in previous 
articles \cite{combes2002,van2003band} are investigated. 
The crystals are square arrays with lattice constant $a$ along the 
$O_x$ and $O_y$ axis, with square or circular rods made 
of a Drude metal and embedded in air (see Fig.\ref{fig:SchemeCrystal}).
As the structure we are dealing with is bi-periodic in the $xy$ plane, 
Bloch-Floquet (or quasi-periodicity) conditions 
\cite{nicolet2004modelling} are imposed on the $x$ and $y$ boundary of 
the square unit cell thanks to the surface 
term of  Eq.\ref{IPPWeakAuxDL}. These quasi-periodic boundary 
conditions are defined by the Bloch wavector $\k = k_x\x+k_y\y$ 
of the electromagnetic eigenmodes. Notice 
that the components $(k_x,k_y)$ of the wavevector 
$\k$ are real while the eigenfrequencies are complex. 
The modal analysis of the structure cell is thus 
performed through the FEM described in Sec.~2.\ref{FEM}, where Bloch 
vector $\k$ is fixed as a parameter spaning the first reduced Brillouin zone.

The dielectric relative permittivity of the rods is given by the single 
Drude resonance:
\begin{equation}
\ep(\om) = 1 - \dfrac{\om_p^2}{\om(\om+\i\gamma)}\, ,
\label{epsDrude}
\end{equation}
with 
\begin{equation}
\dfrac{\om_p a}{2 \pi c} = 1.1 \, , \quad \dfrac{\gamma a}{2 \pi c} = 0.05 \, .
\label{paramDrude}
\end{equation}
The width and radius of the rods have been chosen so that the filling fraction 
of the structures is equal 
to 0.65 in both cases, \textit{i.e.} $w=0.806\,a$ in case of square rods 
and $r = 0.455\,a$ in case of circular rods.
The eigenfrequencies of the structures are retrieved for all wavectors 
$\k$ lying in the first reduced Brillouin zone of the considered crystals 
($k_x \in [0,\pi/a]$ and $k_y \in [0,k_x]$) as represented in Fig.
\ref{fig:SchemeCrystal}.
\subsection{Resonances for s-polarization}
\begin{figure}[h!]
\centering
\fbox{\includegraphics[width=0.97\linewidth]{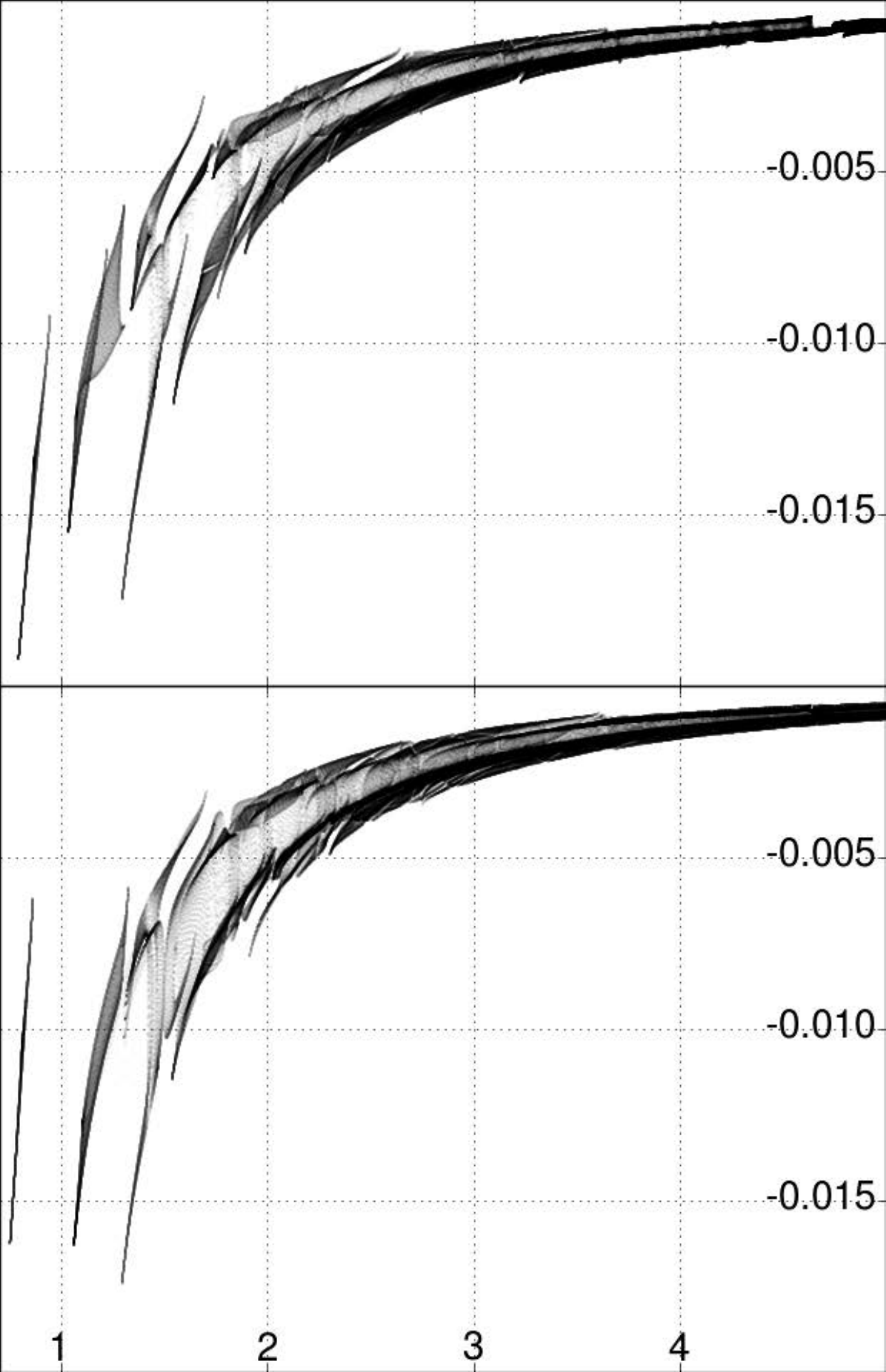}}
\caption{Spectra of the normalized eigenfrequencies $\om_n a /(2 \pi c)$ 
of the photonic crystal of square (top panel) and 
circular (bottom panel) rods for all the wavevectors $\k$ in the first 
Brillouin zone.}
\label{fig:Spectrespol}
\end{figure}

Fig. \ref{fig:Spectrespol} shows the whole spectrum of the photonic crystals made 
of square (top panel) and circular (bottom panel) rods for s-polarization.
The result of this simulation confirms previous results of the 
literature obtained with completely different methods. The exact modal 
method \cite{GDTEM03} is used in Ref.~\cite{combes2002} for square rods, while an expansion on 
cylindrical harmonics combined with lattice sums is used in Ref.~\cite{van2003band} for circular rods. 
Notice that the set of resonances tends to the spectrum of the free Laplacian 
at high frequencies (see Fig. \ref{fig:Spectrespol}) according to the behavior 
of the permittivity in this range ($| \ep(\omega) | \rightarrow 1$ when 
$| \omega | \rightarrow \infty$). 
\begin{figure}[t!]
\centering
\fbox{\includegraphics[width=0.97\linewidth]{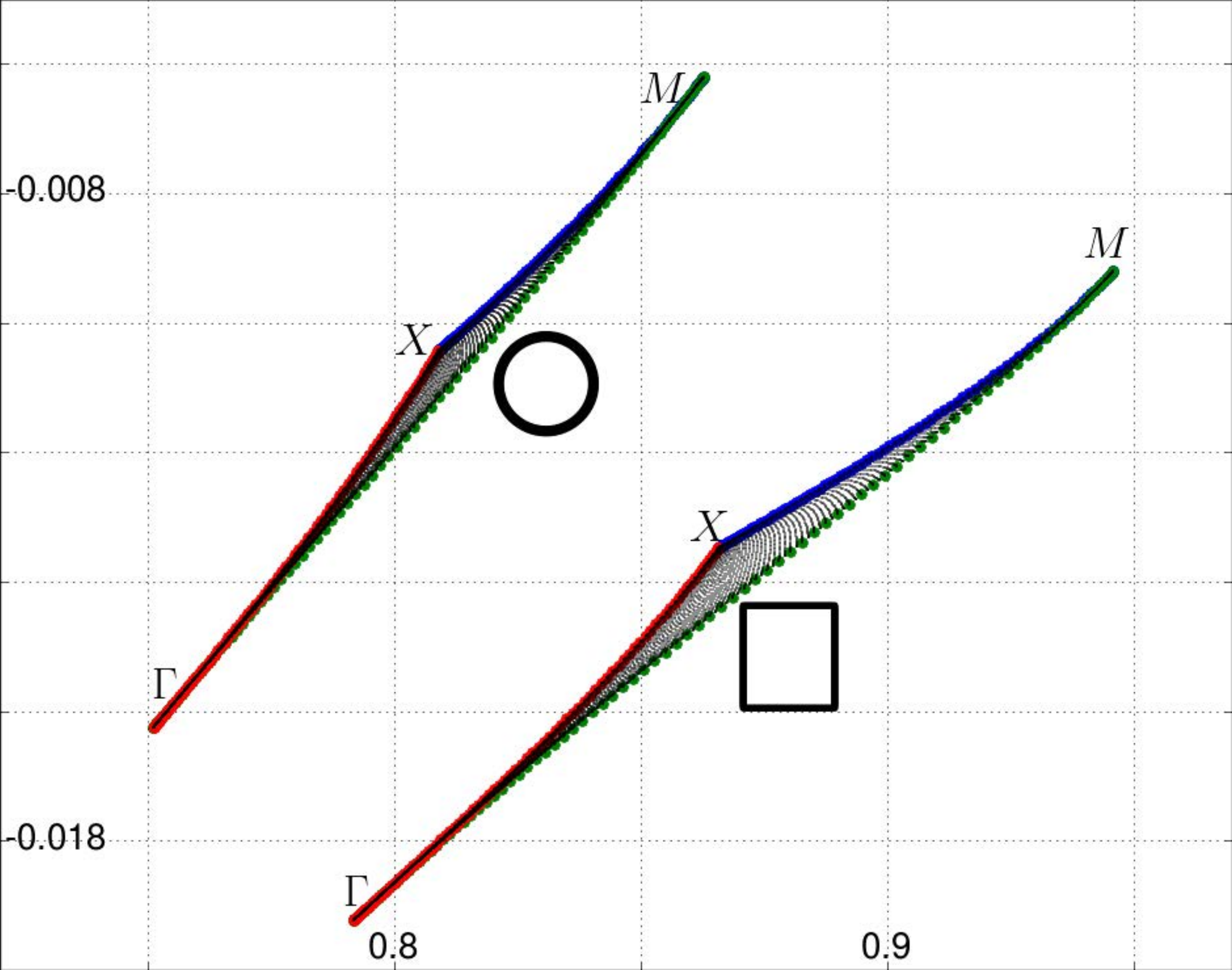}}
\caption{First band of the spectra of the normalized eigenfrequencies 
$\om_n a / (2 \pi c)$.}
\label{fig:Spectre1}
\end{figure}
\begin{figure}[b!]
\centering
\fbox{\includegraphics[width=0.97\linewidth]{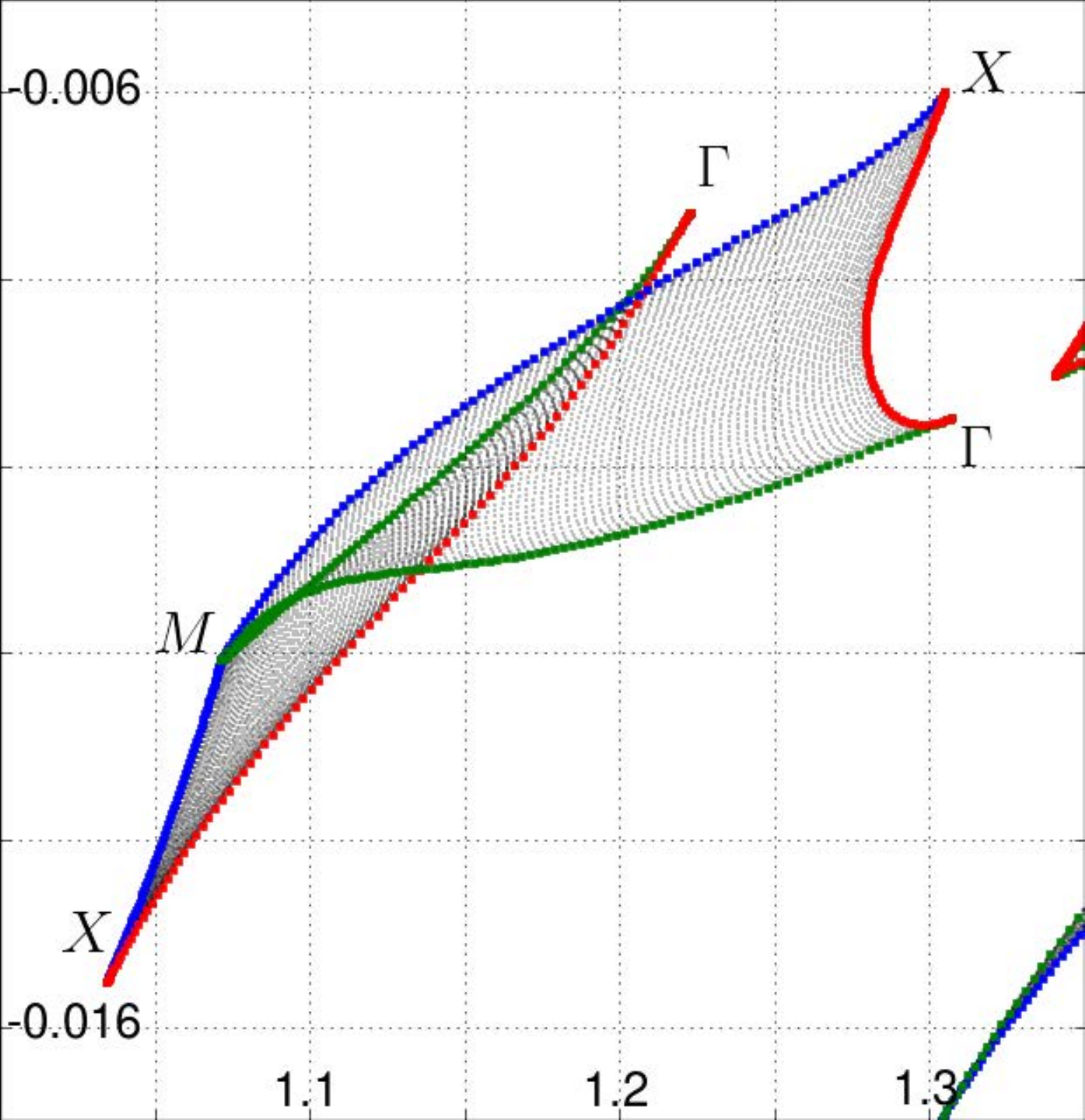}}
\caption{Second and third bands of the spectrum of the normalized resonances 
$\om_n a / (2 \pi c)$. The resonances corresponding to the three boundaries of 
the triangular reduced first Brillouin zone are drawn with colors: 
$\Gamma X$ in red, $\Gamma M$ in green, and $XM$ in  blue. }
\label{fig:Spectre2}
\end{figure}
Also, the complex spectra of the two crystals 
with square and circular rods turn out to be similar, as confirmed by Fig.~\ref{fig:Spectre1} 
showing the first band of complex resonances. Hence, from now on, the investigation 
is focused on the crystal made of square rods.

Fig.~\ref{fig:Spectre2} shows bands 2 and 3 and confirms results presented 
in reference \cite{combes2002} where the complex resonances have been 
rigorously defined and calculated using perturbation theory. 
\begin{figure}[h!]
\centering
\fbox{\hspace*{0.5mm}\includegraphics[width=0.95\linewidth]{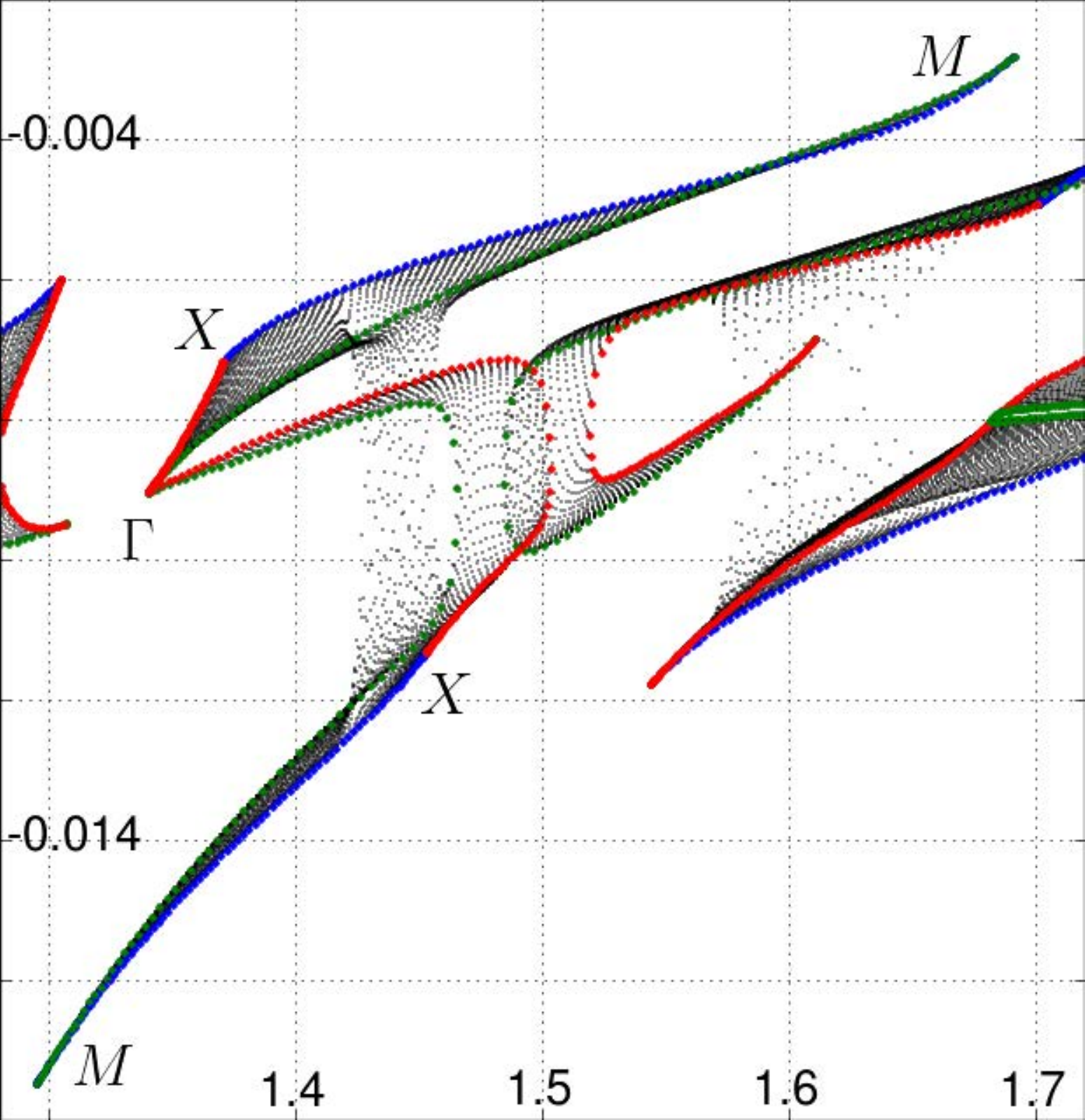}\hspace*{0.5mm}}
\caption{Band 4, band 5 and of higher order of the spectra of the
normalized eigenfrequencies $\om_n a / (2 \pi c)$. The resonances 
corresponding to the three boundaries of the triangular reduced first 
Brillouin zone are drawn with colors: $\Gamma X$ in red, $\Gamma M$ 
in green, and $XM$ in  blue.}
\label{fig:Spectre3}
\end{figure}
Fig.~\ref{fig:Spectre3} shows the bands 4 and 5 (and higher) 
which have even richer structures. 
As pointed out in \cite{van2003band}, calculations of band structures are 
generally performed for wavevector $\k$ spanning the  boundary 
of the first reduced Brillouin zone (in colors on Fig. \ref{fig:SchemeCrystal}), 
the general assumption being that the spectrum of complex resonances is 
located inside the contour formed by the complex resonances corresponding 
to this boundary. Figure \ref{fig:Spectre3} shows a counterexample of this 
assumption where it can be observed that some resonances are located outside 
those deformed triangles, which confirms results obtained in \cite{van2003band} 
for circular rods. 
Moreover, Fig. \ref{fig:Spectre3} reveals that two different bands (bands 4 and 5) 
are connected with a set of eigenfrequencies. Similar behaviors are observed 
for higher order bands. 

In order to have a better understanding of these spectra, the complex 
resonances are represented for wavevector $\k$ along the four ``iso-$k_y$'' lines 
defined by setting the $k_y$ component to the following fixed values: $0.23\pi/a$, 
$0.24\pi/a$, $0.37\pi/a$, and $0.38\pi/a$, while the $k_x$ component spans segments 
$[k_y, 0.5 \pi/a]$). 

\begin{figure}[h]
\centering
\fbox{\hspace*{0.5mm} \includegraphics[width=0.46\linewidth]{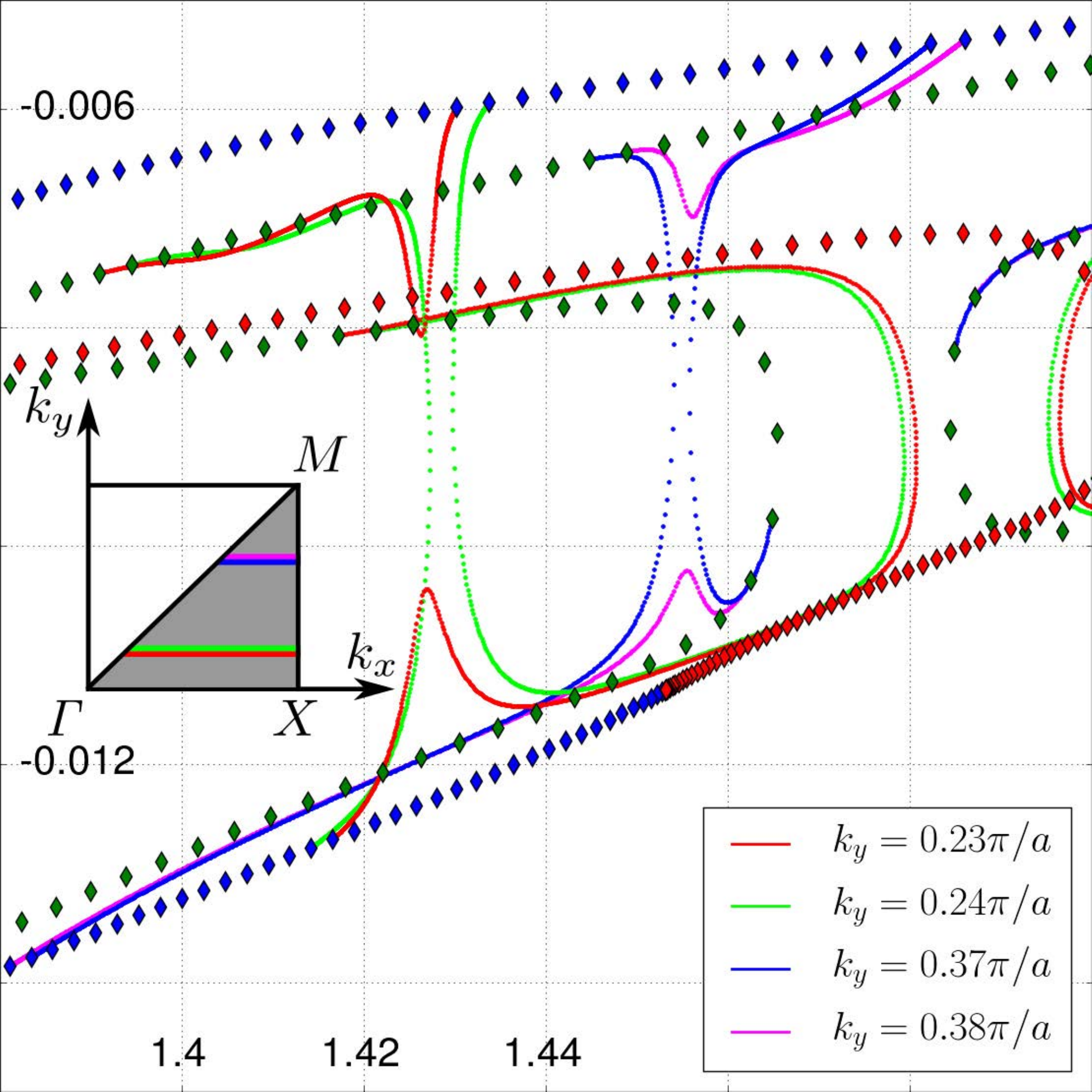}
\hspace*{0.5mm} \includegraphics[width=0.46\linewidth]{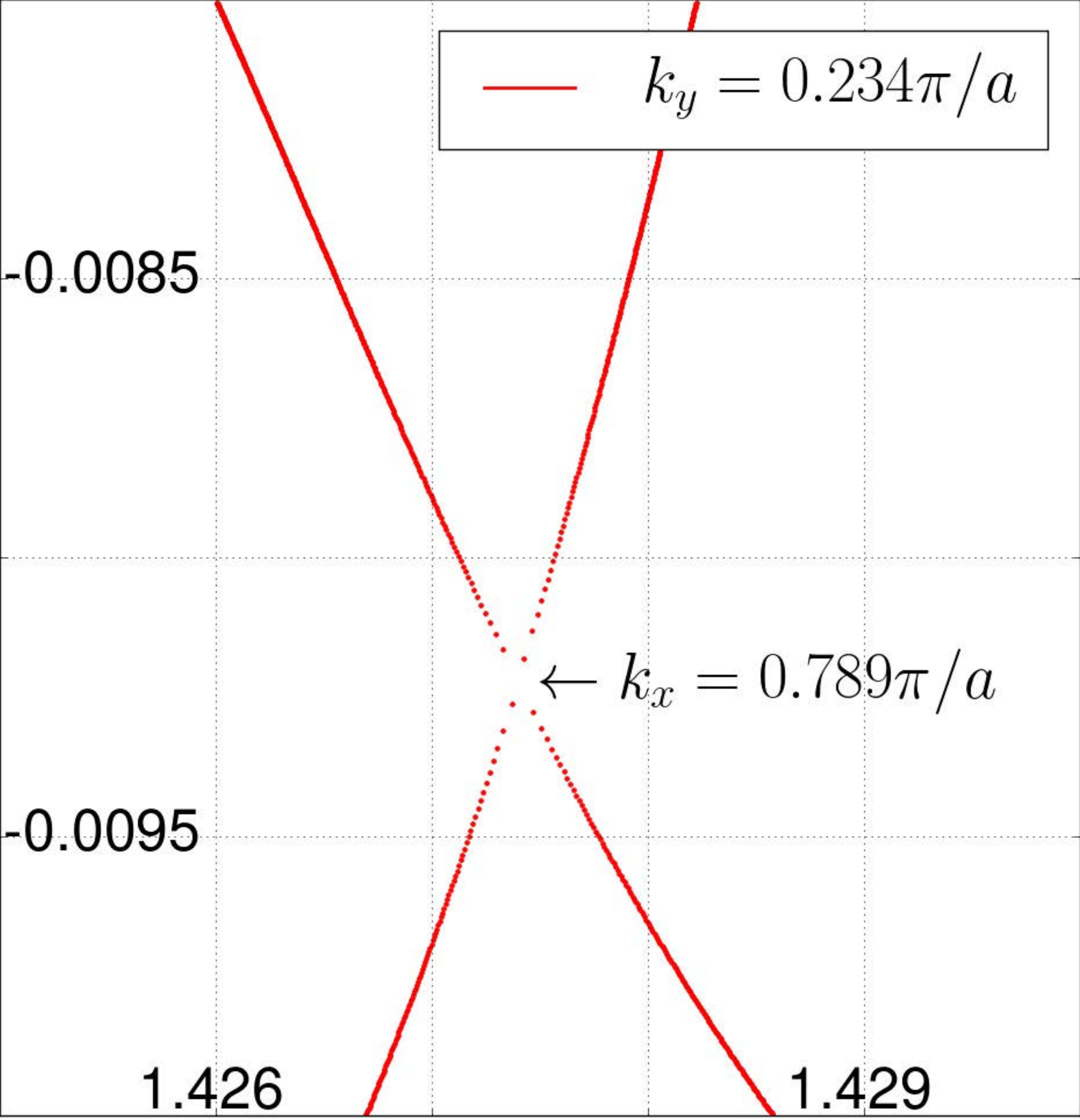} \hspace*{0.5mm}}
\caption{Left panel: Resonances in the photonic crystal made of Drude 
square rods for several iso-$k_y$ lines in bands 4 and 5. Left panel: 
Zoom in the aerea around the critical point $(k_x,k_y) \approx (0.789,0.234) \pi/a$ 
in the band structure.}
\label{fig:Zoom}
\end{figure}
Left panel of Fig. \ref{fig:Zoom} shows iso-$k_y$ lines in the connected 
bands 4 and 5 corresponding to Fig. \ref{fig:Spectre3}. For $k_y = 0.23\pi/a$, 
the path followed by the resonances (red dots) starts from the side  
of the deformed triangle corresponding to $X M$ (blue diamonds), then goes 
out the deformed triangle, and finally comes back to the other side  
of the deformed triangle corresponding to $\Gamma M$ (green diamonds). The part of 
the path leaving the deformed triangle takes the shape of a sharp dip
for the lower absorption band and the shape of a sharp peak for the higher absorption band.
Hence these dip and peak attract seemingly each other. For $k_y = 0.24\pi/a$, 
the path followed by the resonances (green dots) starts from the side  
of the first deformed triangle corresponding to $X M$ (blue diamonds), 
goes out this first deformed triangle to reach the side of the second 
deformed triangle (side corresponding to $\Gamma M$, green diamonds). For this 
value $k_y = 0.24\pi/a$ the two bands are connected. Similar observations 
can be done for values of $k_y$ fixed to $0.38\pi/a$ (violet dots in Fig.~\ref{fig:Zoom} 
showing a dip and a peak) and $0.37\pi/a$ (blue dots 
showing connexion). Therefore, it can be concluded that in a range of 
wavevectors the bands 4 and 5 are enlaced, see Fig. \ref{fig11} 
for a representation of the interlacing. Also, it is stressed that 
the critical wavevector at which the peak and the dip touch each other 
at just one point may be associated to singular properties. 
Right panel of \ref{fig:Zoom} shows the resonances for iso-$k_y$ lines 
around this critical point, whose wavevector has been 
estimated around $(k_x,k_y) \approx (0.789,0.234) \pi/a$. 

\begin{figure}[h]
\centering
\fbox{
\includegraphics[width=0.47\linewidth]{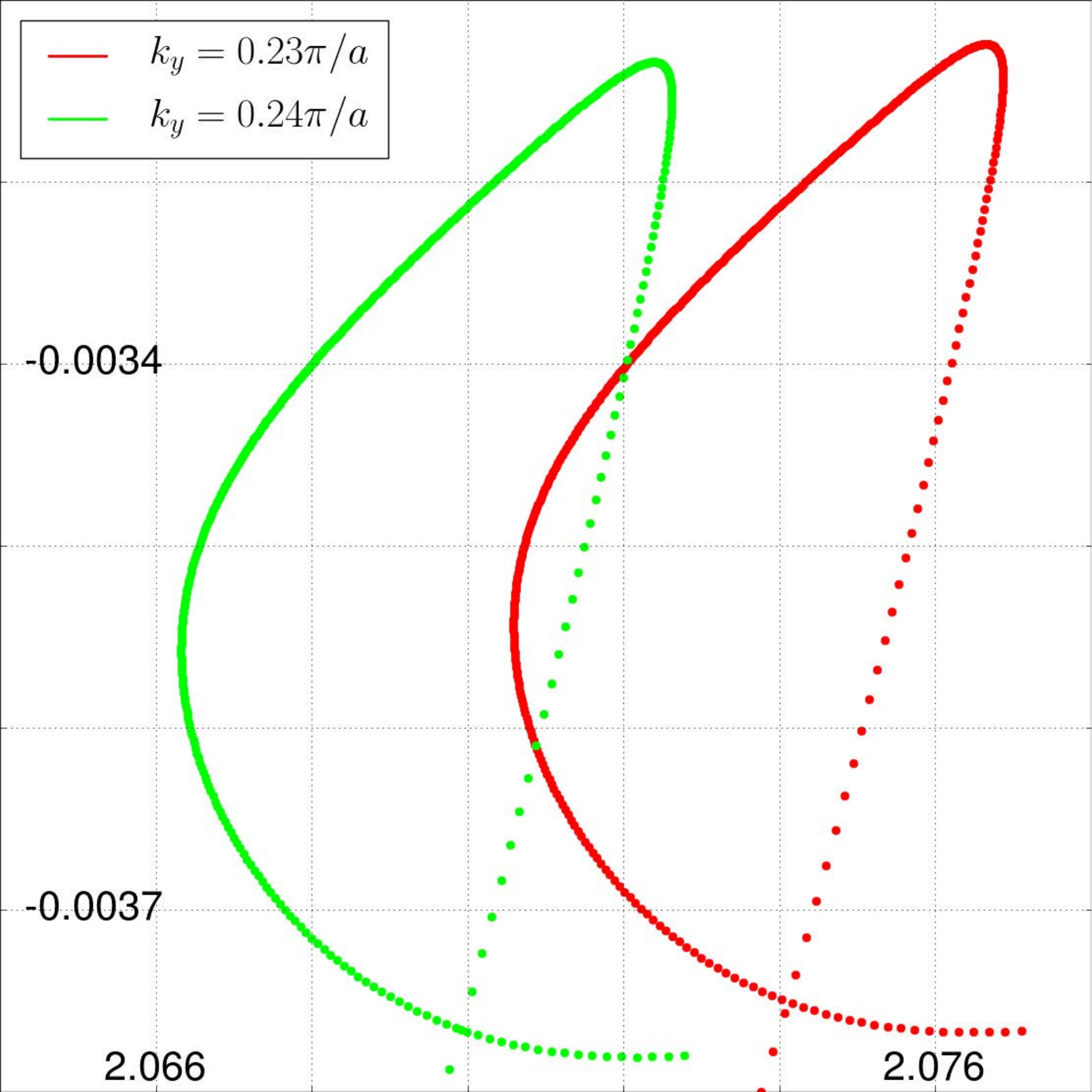}
\includegraphics[width=0.47\linewidth]{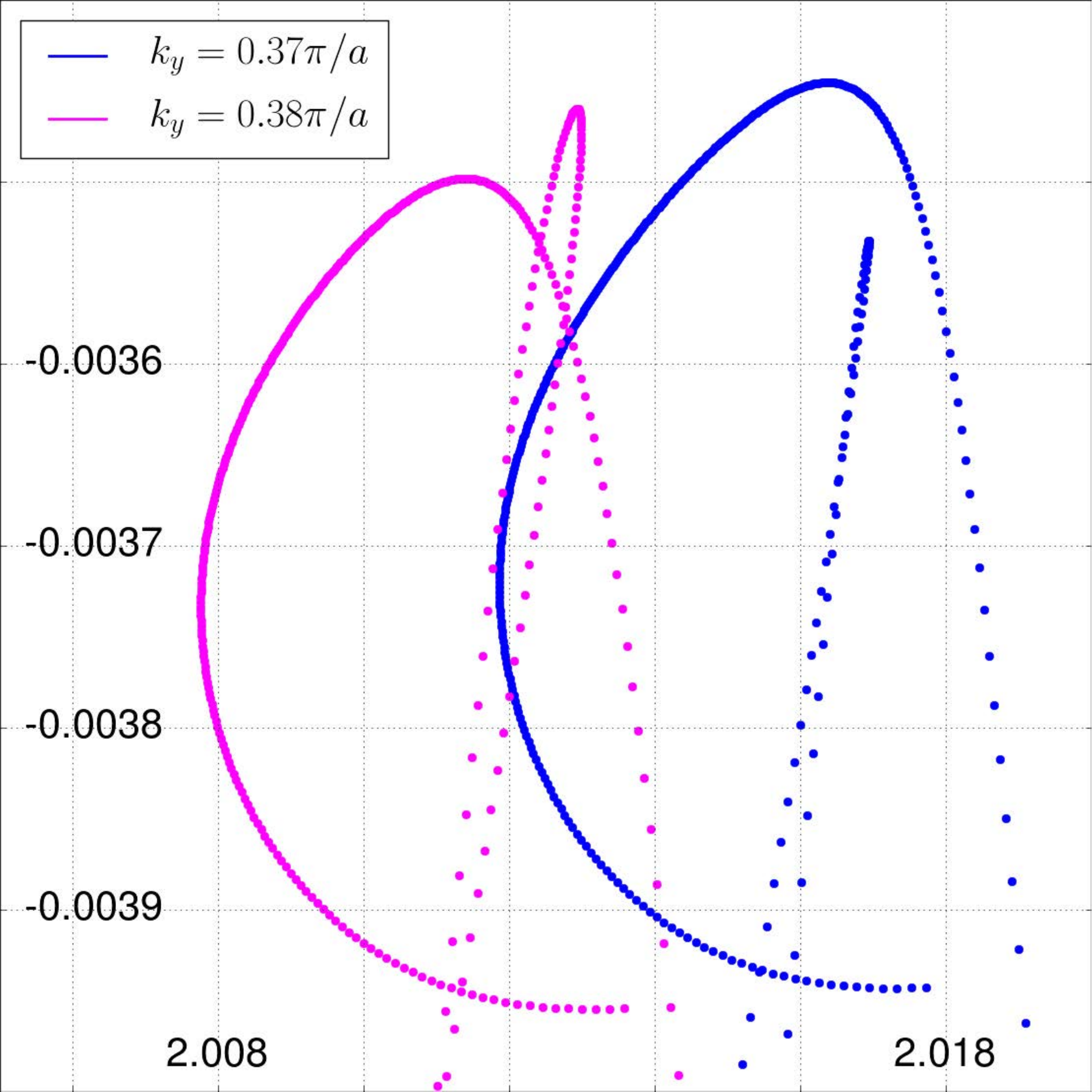}
}
\caption{Resonances in the photonic crystal made of Drude square rods 
for several iso-$k_y$ lines in bands of higher order showing loops 
(left) and cusps (right).}
\label{fig:Zoom2}
\end{figure}

\begin{figure}[h]
\centering
\fbox{
\includegraphics[width=0.42\linewidth]{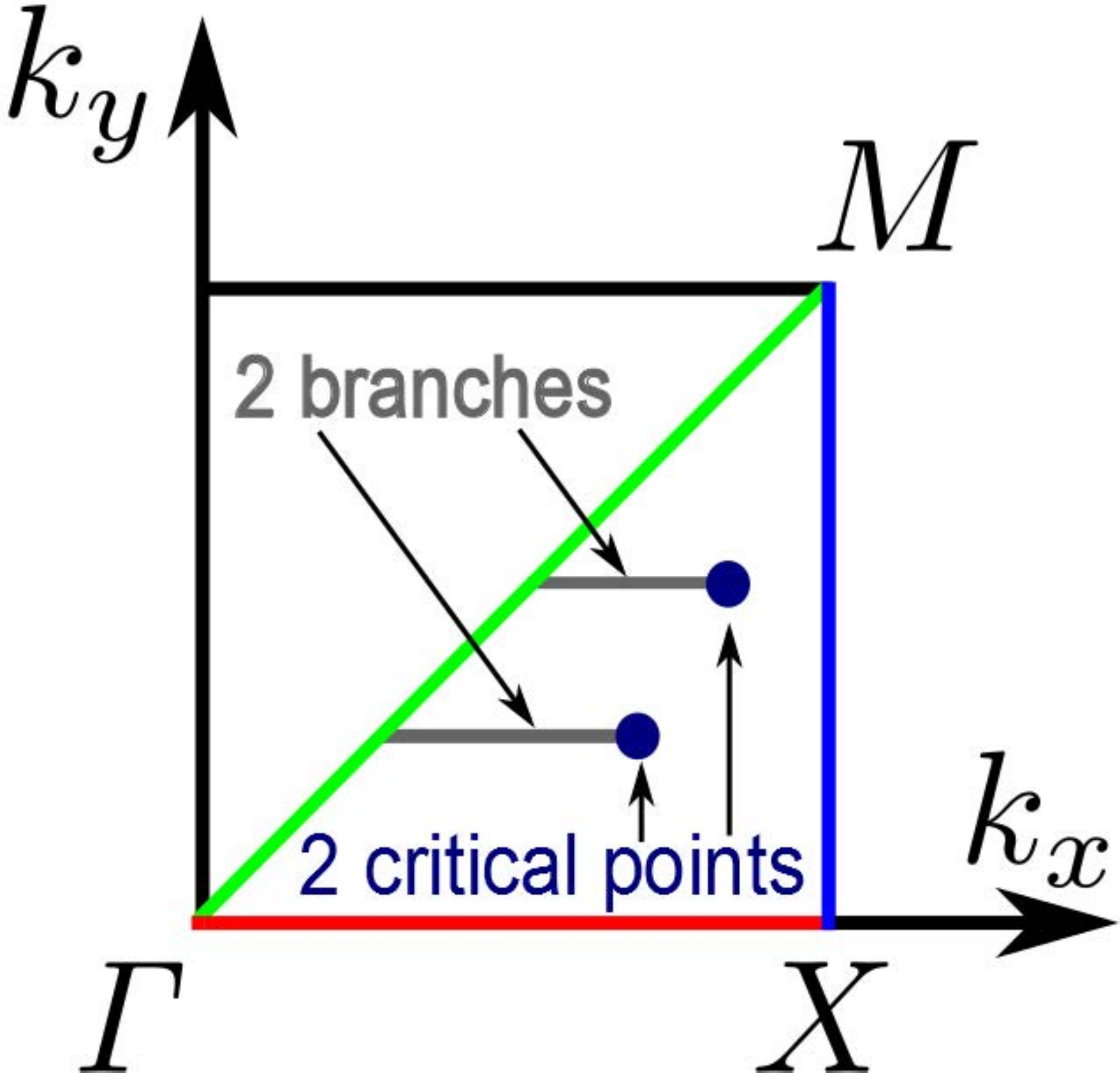}
\hspace*{2mm}
\includegraphics[width=0.49\linewidth]{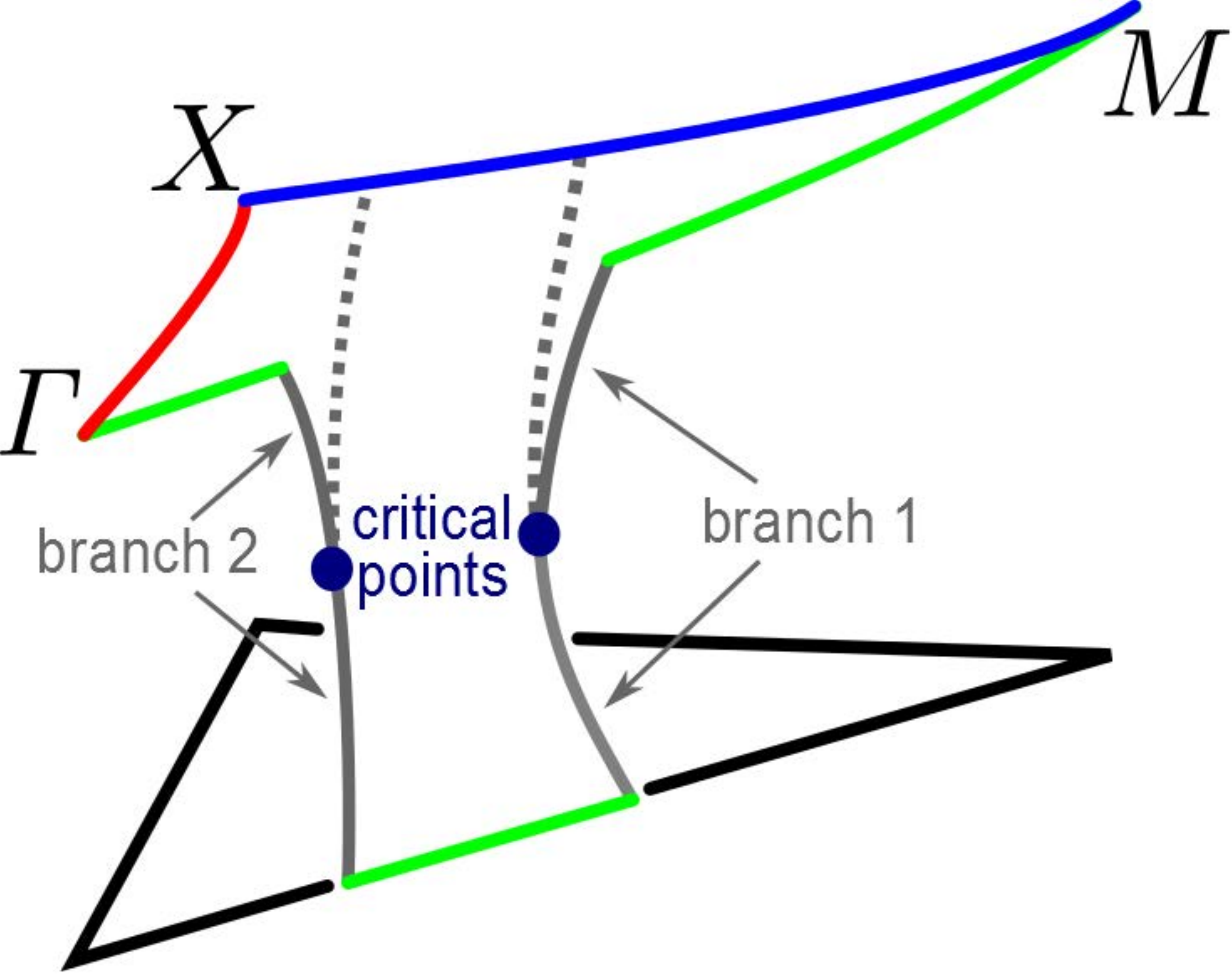}
}
\caption{Representation of the interlacing of the bands}
\label{fig11}
\end{figure}

Fig. \ref{fig:Zoom2} shows iso-$k_y$ lines in bands of higher order 
for $k_y$ fixed to values  
$0.23\pi/a$, $0.24\pi/a$, $0.37\pi/a$, and $0.38\pi/a$. 
The path followed by the resonances exhibits some loops (left panel) and cusps 
(right panel) when $k_x$ spans $[k_y,\pi/a]$. 
These loops and cusps may have important effect on the complex group velocity 
defined as $(\partial\om) / (\partial k_x)$ with changes of direction.

Finally the connection between the two bands 4 and 5 observed in the 
previous figures is discussed. First, it is stressed that the resonances 
corresponding the Brillouin contour (i.e. the contour of the reduced Brillouin zone) 
describe some deformed triangles which do not contain all the resonances 
corresponding to the whole reduced Brillouin zone. Hence the description of the 
solely Brillouin contour does not provide all the information of 
the system: In particular, the critical point highlighted in Fig. \ref{fig:Zoom}
is precisely located in the inner of the reduced Brillouin zone. 
Fig. \ref{fig11} proposes a representation of a band interlaced with a second one, 
as occurs for band 4 with band 5. This representation shows that the set of resonances 
corresponding to the Brillouin contour is not a closed path, 
but a deformed triangle with a cut and a segment. This cut in the deformed triangle 
defines two branches starting from the critical point defined above. Hence
the inside of the Brillouin zone contains two critical points and two branches 
which may have particular properties. This observation supports the relevance 
of a scan of the whole reduced Brillouin zone for some systems. 

Next, it is stressed that the interlacing of bands 4 and 5 may be interpreted 
as a coupling effect. In this case, it can be expected that the coupling of the two 
modes with the same frequency $\om$ and wavevector $\k$ results from a 
symmetry which has been broken. In the present photonic crystal, no spatial 
symmetry is broken. However, in the Drude model (\ref{Drude}), the resonances 
of the permittivity have not the same imaginary part with a first resonance 
at $\om = 0$ and a second at $\om = - i \gamma_D$. Hence, one can imagine that 
this could break some symmetry in the time domain since other models 
for permittivity are generally made by couple of resonances $\pm \om_0 - 
\i \gamma$ (with both $\om_0$ and $\gamma$ real) with the same imaginary part. 
Such assumption is consistent with the interlacing observed 
in Figs. \ref{fig:Spectre3} and \ref{fig:Zoom} which is 
essentially in the vertical (imaginary) direction in order to connect a 
low absorption band to a high absorption band. 

\begin{figure}[h]
\centering
\fbox{\includegraphics[width=0.97\linewidth]{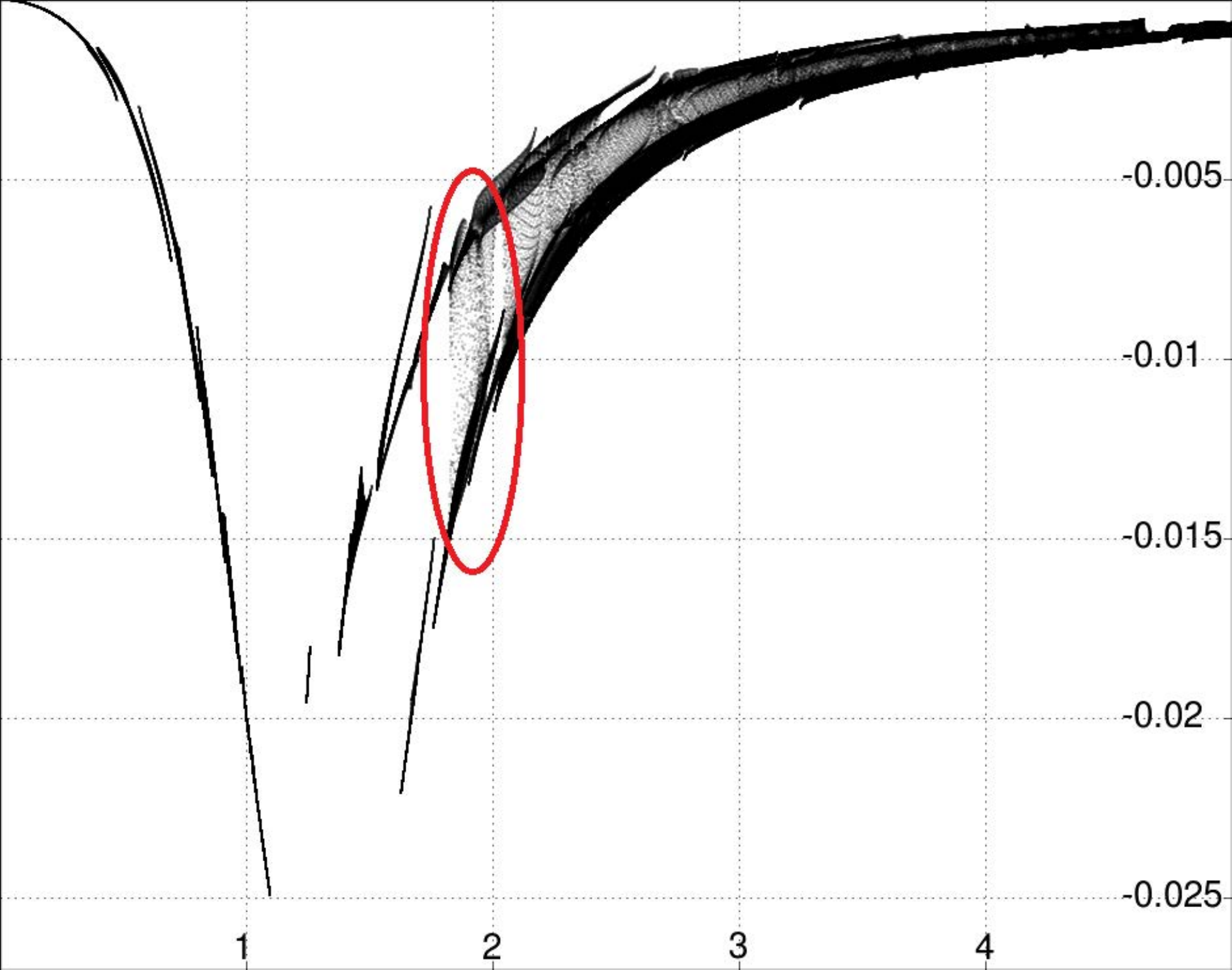}}
\caption{Resonances in the photonic crystal made of Drude-Lorentz square rods. 
The highlighted area shows interlacing of bands. 
}
\label{fig:Lorentz}
\end{figure}

This assumption can been checked by considering the resonances in the same 
crystal with square rods but with the permittivity now given by the 
Drude-Lorentz model (\ref{ep2}) with the parameters
\begin{equation}
\ep_\infty = 1.0 \, , \quad \dfrac{\omega_p a}{2 \pi c} = 1.1 \, , \quad 
\dfrac{\gamma a}{2 \pi c} = 0.05 \, , \quad \dfrac{\omega_0 a}{2 \pi c} = 1.1 \, .
\end{equation}
The corresponding spectrum of resonances represented on Fig. \ref{fig:Lorentz} 
shows that the bands remain enlaced. Hence we conclude that the interlacing of 
bands is not produced by a broken symmetry in this kind of situation: 
further investigations are needed in a future work to understand this 
phenomenon. 
\subsection{Resonances for p-polarization}
\begin{figure}[!b]
\centering
\fbox{\includegraphics[width=0.97\linewidth]{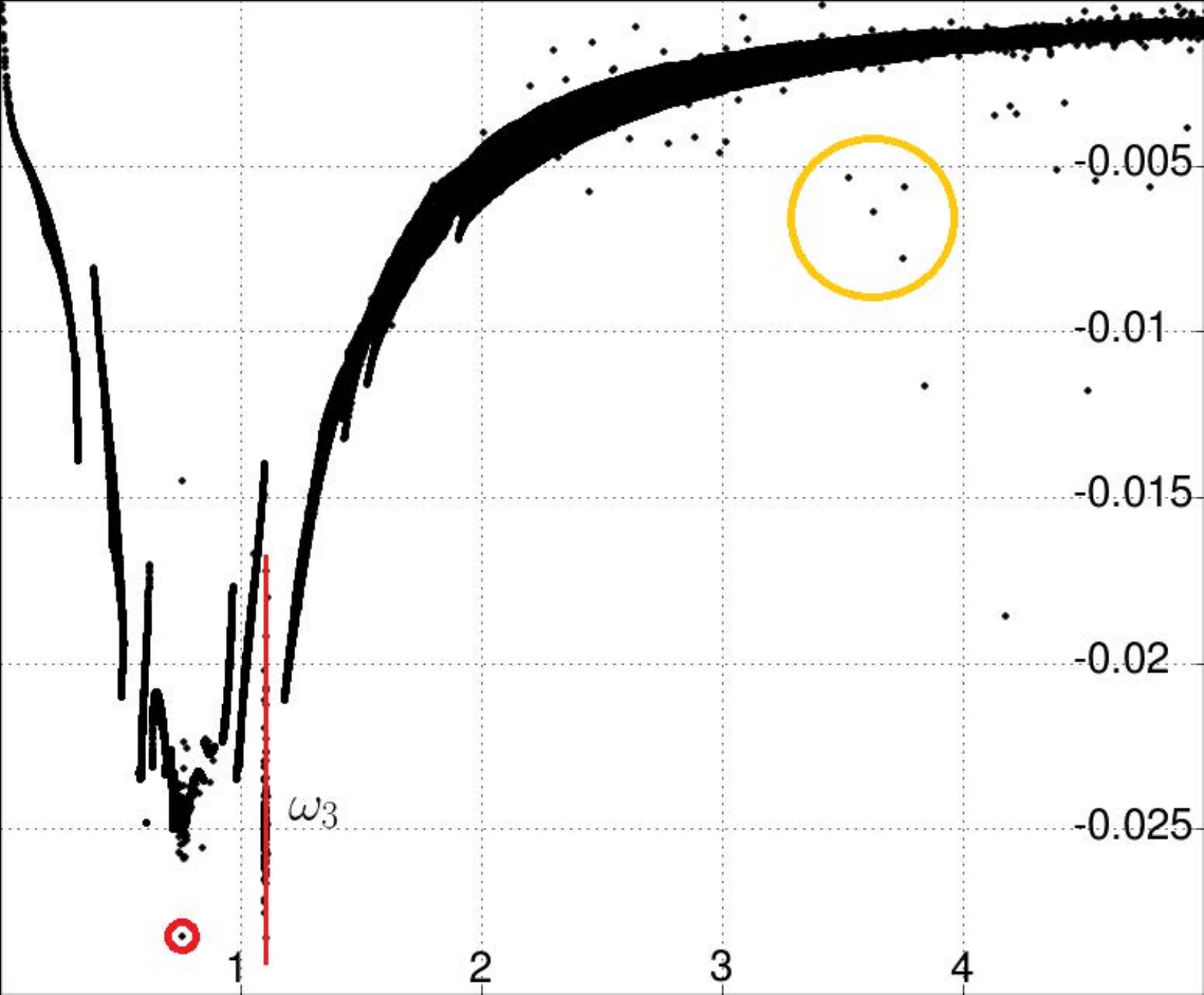}}
\caption{Spectrum of the normalized resonances $\om_n a/(2 \pi c)$
in the photonic crystal made of circular rods of Drude p-polarization. 
Spurious modes produced by $\ep(\om_3) = 0$ are highlighted with red color.
Spurious modes at high frequency are highlighted with yellow color.}
\label{fig:polarPPC}
\end{figure}

The same numerical calculations have been performed for p-polarization 
with Drude parameters given by (\ref{epsDrude}) and (\ref{paramDrude}). 
In this case only the photonic crystal of circular rods is considered 
in order to compare the results to the one of the literature \cite{van2003band}. 
Fig. \ref{fig:polarPPC} presents the whole spectrum of complex resonances.

The results are in agreement with the ones in the literature \cite{van2003band}, except 
for the presence of spurious modes as previously observed for closed cavities in Sec.~3. 
These spurious modes affect the results around the complex frequency $\om_3 \simeq 1,1 - 0.025\i$ 
where $\ep(\om_3) = 0$. Notice that these spurious modes are mostly located 
on a vertical axis starting from the vanishing frequency $\om_3$ (red line on Fig. 
\ref{fig:polarPPC}). In addition, the presence of spurious modes is observed at high 
frequencies (yellow circle on Fig. \ref{fig:polarPPC}). The presence of these additional modes 
can be explained by the difficulty to describe the high frequency regime, especially in this 
p-polarization case which more demanding for the coarse mesh and the choice of edge elements. 

This preliminary implementation in p-polarization needs further development which will 
be presented in a future work. 
\section{Conclusion}
In this paper, the so-called ``resonance formalism'' has been implemented
into the Finite Element Method in order to perform modal analysis of dispersive and 
dissipative photonic structures. The method has been first validated in the simple case 
of a closed cavity and then used in the case of two-dimensional crystals. The 
numerical results obtained in s-polarization are in excellent 
agreement with previous results in the literature and obtained with completely different 
numerical methods. As to results obtained in p-polarization, they suffer from the presence 
of spurious modes which, in the case of the used FEM method, can be explained by the 
non-fulfillment of the divergence condition when the permittivity vanishes. 

For s-polarization, the numerical method presented in this paper provides an efficient tool 
to compute and analyze the complex spectrum of dispersive systems and non self-adjoint operators 
which have been rarely explored. The present results confirm that retrieving the 
resonances for wavevectors corresponding to the boundary of the first reduced Brillouin zone is 
not sufficient to catch the whole spectrum of photonic crystals. In particular, the presence of 
critical points and branches have been highlighted in the inside of the reduced Brillouin zone. 
Also, it has been shown that the phenomenon of resonances outgoing of the contour is associated 
to an interlacing of bands. However, 
no argument has been found to explain this interlacing of bands, particularly a coupling of 
modes could not be evidenced. 
In addition to this band interlacing, cusps and loops have been highlighted in the band 
structure, which could have interesting effect on the group velocity. 

Further developments of the present method are required to be applied on
open-systems and 3D structures. Regarding open-systems, the solution is to use Perfectly 
Matched Layer damping the field in free space \cite{berenger1994perfectly}. For 3D systems, 
a solution to avoid spurious modes while using edge elements in the FEM has to be developed.

\section*{Funding Information}
This work is supported by the Agence Nationale de la Recherche (project PLANISSIMO ANR-12-NANO-0003).
\section*{Acknowledgment}
Philippe Lalanne (CNRS, LP2N, France) is acknowledged for fruitfull discussions. 
\bigskip



\section{Appendix}
\subsection{Extension to the absorptive case 
of the purely dispersive auxiliary field formalism.}
\label{AnnexAuxD}
A material with relative permittivity described by a single Drude-Lorentz 
resonance is considered. With
$\omg = \sqrt{\omD^2-\gammaD^2/4}$ and $\ompm = \pm \omg - \i \gammaD/2$, 
the expression (\ref{DrudeLorentz})  becomes
\begin{equation}
\begin{array}{ll}
\ep(\om) &\hspace{-2mm} = \epinf - \dfrac{\om_p^2}{\om^2+\i\,\gammaD\,\om - \omD^2} = 
\epinf - \dfrac{\om_p^2}{(\om - \om_+) (\om - \om_-)}\\[2mm]
&\hspace{-2mm}= \epinf - \dfrac{\om_p^2}{2\omg}\left[\dfrac{1}{\om-\om_+}-\dfrac{1}{\om - \om_-}\right]\,.
\end{array}
\label{DrudeLorentz}
\end{equation}
The corresponding dielectric susceptibility $\chi$ can be retrieved 
through the inverse Fourier transform of $\ep(\om) - \ep_\infty$:
\begin{equation}
\chi(t) = -\dfrac{1}{2\pi}\dint_{\R} d\om \dfrac{\exp(-\i\om t)\om_p^2}{[\om-\om_+][\om-\om_-]}\,.
\label{ChiDL}
\end{equation}
For negative time, this integral is computed by closing the loop in the upper half plane of complex 
frequencies $\om$. Since the function under the integral in Eq. (\ref{ChiDL}) has no pole in the upper 
half plane, the dielectric susceptibility vanishes according to the causality principle.
For positive time, the integral is now computed by closing the loop in the lower half plane of 
frequencies $\om$ with negative imaginary part. The function under the integral has now two 
poles in the lower half plane at $\om_\pm$. Using the Cauchy residue theorem, the dielectric 
susceptibility is obtained:
\begin{equation}
\begin{array}{ll}
\chi(t)& \hspace{-2mm}= \dfrac{\om_p^2}{2i \omg} 
\big\{ \exp[-\i \om_-t]-\exp[ - \i \om_+t] \big\} \\[2mm]
& \hspace{-2mm} = \dfrac{\om_p^2}{\omg} \, \exp[-t\gammaD/2] \, \sin[\omg t] \, .
\end{array}
\label{ChieDL}
\end{equation}
Maxwell's equations without sources in the time regime for the considered material are written:
\begin{equation}
\begin{array}{ll}
\rot\E(\r,t) & \hspace{-2mm} = -\muv \dfrac{\partial \H}{\partial t}(\r,t)\\[2mm]
\rot\H(\r,t )& \hspace{-2mm} =  \dfrac{\partial \D}{\partial t}(\r,t) = 
\dfrac{\partial\left[\epv\epinf\E(\r,t) + \P(\r,t)\right]}{\partial t}\, ,
\end{array}
\label{MaxAux}
\end{equation}
where $\D$ and $\P$ are the displacement and polarization fields.
Noticed that $\epinf$ in Eq. (\ref{MaxAux}) is considered as a constant 
relative permittivity acting directly on the electric field, and not a 
part of the polarization field. This choice has been made since the expression of 
the polarization requires a time convolution for a rigorous formulation of 
the problem. In real world this trick does not appear since
the constant $\epinf$ must take the unit value. However, this additional parameter 
is introduced because it allows efficient fit of the real permittivity of metals
\cite{barchiesi2014errata}. 

The polarization field is given by
\begin{equation}
\P(\r,t) = \ep_0\dint_{-\infty}^{t} ds \, \chi(\r,t-s) \E(\r,s) \, .
\label{P}
\end{equation}
Using (\ref{ChieDL}), the expression of the time derivative of the polarization 
field (also called the microscopic currents) is obtained:
\begin{equation}
\begin{array}{ll}
\dfrac{\partial \P}{\partial t}(\r,t) & \hspace{-2mm} = 
\epv\dfrac{\partial}{\partial t}\dint_{-\infty}^{t} ds \, \chi(\r,t-s) \E(\r,s)\\[2mm]
& \hspace{-2mm} = \epv \chi(\r,0) \E(\r,t) + \epv 
\dint_{-\infty}^{t} ds \, \dfrac{\partial\chi}{\partial t}(\r,t-s) \E(\r,s)\\[2mm]
& \hspace{-2mm} = \epv \dfrac{\om_p^2}{2\omg} \, \om_+ 
\dint_{-\infty}^{t} ds \,\exp[ - \i\om_+(t-s)] \E(\r,s) \\[2mm]
& \hspace{-2mm} - \epv \dfrac{\om_p^2}{2\omg} \, \om_- 
\dint_{-\infty}^{t} ds \,\exp[ - \i\om_-(t-s)] \E(\r,s) \, . \\[2mm]
\end{array}
\label{Px}
\end{equation}
Hence the two auxiliary fields defined below appear naturally
\begin{equation}
\Ae^\pm(\r,t) = \mp \i \dfrac{\om_p}{\sqrt{2} \omg} \, \om_\pm 
\dint_{-\infty}^{t}ds \, \exp[- \i \om_\pm (t-s)]\E(\r,s) \, ,
\label{AuxDLAnnex}
\end{equation}
and the time derivative of the polarization field becomes
\begin{equation}
\dfrac{\partial \P}{\partial t}(\r,t) = \i \epv \dfrac{\om_p}{\sqrt{2}} \Ae^+(\r,t) 
+ \i \epv \dfrac{\om_p}{\sqrt{2}} \Ae^-(\r,t) \, .
\label{dtPx}
\end{equation}
Finally, Maxwell's equations result in the set of equations (\ref{TempMaxDL}), 
denominated as the ``resonance formalism''.

For a Drude resonance, similar arguments lead to the definition of the 
auxiliary field Eq. (\ref{AuxD}) and Maxwell's equations become (\ref{TempMaxD}).

\subsection{Eigenfrequencies of closed dispersive cavities.}
\label{EigenCavity}
The expressions (\ref{TransS}) and (\ref{TransP}) of the dispersion law are 
established for the system considered in section 3, and which is made of two square 
domains (side length $a$) of permittivity $\ep_1(\om)$ and $\ep_2(\om)$.
\subsubsection{s-polarization.}
In this case, the electric field has a single component $E(x,y,\om)$ along the $z$-axis 
and Dirichlet boundary conditions are imposed on the boundary of the domain.
The following eigenvalue problem has to be solve:
\begin{equation}
\left[ \dvxs + \dvys + \om_n^2\ep(\om_n)\epv\muv\right]E_n(x,y) = 0 
\label{PbS}
\end{equation}
Using the separation of variables, the eigenfunctions $E_n$ are determined 
with the decomposition
\begin{equation}
E_n(x,y) = \phi_p(x) \psi_q(y)\, .
\label{varsep}
\end{equation}
The solution in the $y$ direction is just 
\begin{equation}
\psi_q(y) = \sin[q \pi y / a] \, , \quad q \in \mathbb{N} \setminus \{0 \} \, , 
\label{psiq}
\end{equation}
which ensures the Dirichlet boundary conditions. This solution is 
injected in the equation (\ref{PbS}):
\begin{equation}
\dfrac{d^2 \phi_p}{dx^2}(x) + \left[\om_n^2\ep(\om_n)\epv \muv - q^2\pi^2/a^2 \right] \phi_p(x) = 0 \, .
\label{Psi}
\end{equation}
The propagation constants are defined as 
\begin{equation}
\beta_j(\om_n) = \sqrt{\dfrac{\om_n^2}{c^2}\ep_1(\om_n)-\dfrac{q^2\pi^2}{a^2}} \, , \quad j=1,2 ,
\label{beta-an}
\end{equation}
and the equation (\ref{Psi}) becomes 
\begin{equation}
\begin{cases}
[d^2\phi_p/dx^2 ](x) + \beta_1^2(\om_n)\phi_p(x) =0 \hspace*{10mm} \text{ if } x<0\\[2mm]
[d^2\phi_p/dx^2 ] (x) + \beta_2^2(\om_n)\phi_p(x) =0 \hspace*{10mm} \text{ if } x>0
\end{cases}\, .
\end{equation}
The general solution of these equations are
\begin{equation}
\begin{cases}
\phi_p(x) = A_1 \cos[\beta_1(\om_n)x]+B_1\sin[\beta_1(\om_n)x] \hspace*{10mm} \text{ if } x<0\\[2mm]
\phi_p(x) = A_2\cos[\beta_2(\om_n)x]+B_2\sin[\beta_2(\om_n)x] \hspace*{10mm} \text{ if } x>0
\end{cases}\, .
\label{system}
\end{equation}
The four unknowns are determined using the four boundary conditions: $\phi_p(-a) = 0$, 
$\phi_p(a) = 0$, $\phi_p(0^-) = \phi_p(0^+)$ and $[d\phi_p/dx](0^-) = [d\phi_p/dx](0^+)$.
This leads to the following set of equations
\begin{equation}
\begin{cases}
A_1\cos[\beta_1(\om_n)a]=B_1\sin[\beta_1(\om_n)a]\\[2mm]
A_2\cos[\beta_2(\om_n)a]=-B_2\sin[\beta_2(\om_n)a]\\[2mm]
A_1 = A_2\\[2mm]
B_1\beta_1(\om_n) = B_2\beta_2(\om_n)
\end{cases}\, .
\end{equation}
The existence of a non vanishing solution to this set of equations provides 
the dispersion equation of the system:
\begin{equation}
\dfrac{\tan[\beta_1(\om_n)a]}{\beta_1(\om_n)}+\dfrac{\tan[\beta_2(\om_n)a]}{\beta_2(\om_n)} = 0\, .
\label{vpE}
\end{equation}
\subsubsection{p-polarization.}
In this case, the magnetic field has a single component $H(x,y,\om)$ along the $z$-axis 
and Neumann boundary conditions are imposed on the boundary of the domain.

The different steps presented in the s polarization case can be used. The first difference 
is the expression of the solution in the $y$ direction 
\begin{equation}
\psi_q(y) = \cos[q \pi y / a] \, , \quad q \in \mathbb{N} \, , 
\label{psiq}
\end{equation}
which now ensures the Neumann boundary conditions. And the second difference is the 
boundary conditions used to solve the system (\ref{system}) which are now: 
$[d\phi_p/dx](-a) = 0$, $[d\phi_p/dx](a) = 0$, $\phi_p(0^-) = \phi_p(0^+)$ and 
$\ep_2(\om_n)[d\phi_p/dx](0^-) = \ep_1(\om_n)[d\phi_p/dx](0^+)$. The resulting 
linear set of equations is
\begin{equation}
\begin{cases}
A_1\sin[\beta_1(\om_n)a]=- B_1\cos[\beta_1(\om_n)a]\\[2mm]
A_2\sin[\beta_2(\om_n)a]=B_2\cos[\beta_2(\om_n)a]\\[2mm]
A_1 = A_2\\[2mm]
B_1\beta_1(\om_n)/\ep_1(\om_n) = B_2\beta_2(\om_n)/\ep_2(\om_n)
\end{cases}\, .
\end{equation}
Again, the existence of a non vanishing solution to this set of equations provides 
the dispersion equation of the system:
\begin{equation}
\dfrac{\beta_1(\om_n)}{\ep_1(\om_n)}\tan[\beta_1(\om_n)a]+
\dfrac{\beta_2(\om_n)}{\ep_2(\om_n)}\tan[\beta_2(\om_n)a] = 0\, .
\label{vpH}
\end{equation}

\end{document}